\documentclass[aps,prd,twocolumn,floatfix,nofootinbib,superscriptaddress]{revtex4-2}
\usepackage{amsmath,amssymb,bm,mathrsfs}
\usepackage{newtxmath,newtxtext}
\pdfmapfile{+newtx.map}
\usepackage{graphicx}
\usepackage{textcomp}
\usepackage{booktabs}
\usepackage{dcolumn}
\usepackage{ragged2e}
\usepackage{hyperref}
\usepackage{xcolor}
\usepackage{epsfig}
\usepackage{appendix}
\usepackage{orcidlink}
\usepackage{multirow}
\usepackage{tcolorbox}
\usepackage[section]{placeins}
\makeatletter
\long\def\@makecaption#1#2{%
  \par
  \vskip\abovecaptionskip
  \begingroup
    \small\rmfamily
    \noindent
    \@make@capt@title{#1}{#2}\par
  \endgroup
  \vskip\belowcaptionskip
}
\makeatother
\hypersetup{colorlinks=true, linkcolor=black, urlcolor=black, citecolor=black}
\hypersetup{hypertexnames=false}

\colorlet{RED}{red}
\newcommand{\calR}{{\cal R}}
\newcommand{\calP}{{\cal P}}

\newcommand{\dd}{\mathrm{d}}

\newcommand{\bq}{\begin{equation}}
\newcommand{\eq}{\end{equation}}
\newcommand{\bqn}{\begin{eqnarray}}
\newcommand{\eqn}{\end{eqnarray}}
\newcommand{\nb}{\nonumber}
\newcommand{\lb}{\label}

\begin{document}

\title{Primordial Non-Gaussianity and Correlation-Function Perturbativity in mLQC-I: The Hybrid Approach}

\author{Rui Pan}
\email{rui\_pan1@baylor.edu}

\author{Jamal Saeed}
\email{Jamal\_Saeed1@baylor.edu}

\author{Anzhong Wang}
\email{anzhong\_wang@baylor.edu}
\thanks{Corresponding author}

\affiliation{GCAP-CASPER, Department of Physics and Astronomy, Baylor University, Waco, TX 76798-7316, USA}

\date{November 11, 2026}

\begin{abstract}

We investigate primordial non-Gaussianity and the perturbative consistency of scalar cosmological
perturbations in modified loop quantum cosmology I (mLQC-I) within the hybrid approach. We construct
the cubic interaction Hamiltonian in the same canonical representation as the quadratic theory and
derive an exact finite-momentum in-in expression for the tree-level bispectrum. We consider two
initial-state prescriptions: a Gaussian state specified at the bounce and the de Sitter vacuum specified at a remote contracting surface. The retained nonlinear-conversion correction is assessed using
$\Xi=|\Delta{\cal P}_{\cal R}/{\cal P}_{\cal R}^{(0)}|$, where ${\cal P}_{\cal R}^{(0)}$ is the
leading-order curvature power spectrum and $\Delta{\cal P}_{\cal R}$ is the retained nonlinear-conversion
correction. For the bounce-defined initial state, a conservative, cancellation-independent,
finite-window absolute-integrand estimate is below $6\times10^{-6}$ at each of the five sampled
momenta, whereas for the remote de Sitter initial state the signed correction is dominated by
cancellations and decreases from $\Xi\simeq9.9\times10^{-2}$ at one tenth of the bounce-curvature
scale to $\Xi\simeq2.3\times10^{-4}$ at the bounce-curvature scale. Although the retained two-point
corrections remain below unity, the primordial bispectrum can have very large pointwise amplitudes:
the bounce-defined state yields an equilateral $f_{\rm NL}$ with a root-mean-square amplitude
over the oscillation phase of order $10^3$--$10^4$, while the remote state can yield much larger
values. The dominant rapid oscillation is organized approximately by $k_t=k_1+k_2+k_3$, with
$\Delta k_t\simeq5.7\times10^{-4}m_{\rm P}$, while the triangle-shape envelope varies slowly.
For the illustrative bounce-to-present mapping adopted here, CMB-projection diagnostics indicate
substantial phase averaging. Thus, large pointwise primordial non-Gaussianity can coexist with
a small retained correction to the two-point function. 
%The projected residuals are illustrative
%and require further numerical convergence and domain checks.

\label{v5removed:1}

\end{abstract}

\maketitle

\section{Introduction}
\renewcommand{\theequation}{1.\arabic{equation}}
\setcounter{equation}{0}
\label{sec:intro}

Loop quantum cosmology (LQC) replaces the classical big-bang singularity by
a quantum bounce and provides a setting in which cosmological perturbations
can be propagated through a Planck-curvature epoch
\cite{Ashtekar:2011ni,Agullo:2023rqq}.  Two widely used frameworks for
cosmological perturbations are the dressed-metric and hybrid approaches.
In both cases the homogeneous background is loop quantized, while the
inhomogeneous perturbations are quantized in a Fock representation.  In the
dressed-metric approach, perturbations propagate as quantum fields on an
effective geometry dressed by the quantum background
\cite{Agullo:2012sh,Agullo:2012fc,Agullo:2013ai}, whereas in the hybrid
approach the truncated background-plus-perturbation system is quantized as
a constrained system, combining polymeric quantization of the homogeneous
sector with Fock quantization of the perturbations
\cite{Fernandez-Mendez:2012poe,Fernandez-Mendez:2013jqa,
Fernandez-Mendez:2014raa}.  Although the two formulations are related at
the classical level, their effective perturbation theories need not coincide
after the background structures entering the perturbative Hamiltonian are
regularized \cite{Li:2022evi,Iteanu:2022zha}.

Modified loop quantum cosmologies arise from regularization ambiguities in
the gravitational Hamiltonian constraint.  In standard LQC the Euclidean
and Lorentzian contributions are first related classically in spatially flat
FLRW spacetime and the resulting combined expression is then quantized.
An alternative route is to quantize these two contributions separately,
more closely following their treatment in full loop quantum gravity.  This
possibility was first investigated by Yang, Ding, and Ma
\cite{Yang:2009fp}, leading to the two quantizations now known as mLQC-I
and mLQC-II.  The quantum dynamics of mLQC-I in the improved
$\bar\mu$ scheme was subsequently developed in
Ref.~\cite{Assanioussi:2018hee}.  A related effective Hamiltonian also
emerged from a top-down construction based on full loop quantum gravity
\cite{Dapor:2017rwv}, and the improved $\bar\mu$ dynamics was later recovered
when graph-changing effects were incorporated \cite{Han:2021cwb}.  The
effective background dynamics of mLQC-I and mLQC-II were analyzed
systematically in Refs.~\cite{Li:2018opr,Li:2018fco}; see also
Ref.~\cite{Li:2021mop} for a review.

A particularly important feature of mLQC-I is the intrinsic asymmetry of its
quantum bounce.  The expanding branch approaches classical general
relativity at low curvature, whereas the remote contracting phase approaches
a quantum-geometrical de Sitter spacetime \footnote{Recently, in the Bianchi~I model, a quantum mechanism of shear
damping was proposed, which can lead to a complementary branch structure.
In this scenario, the pre-bounce branch is asymptotically described by
classical general relativity, whereas the post-bounce branch approaches a
quantum inflationary phase characterized by an effective Planck-scale
cosmological constant \cite{Gan:2025uvt,Gan:2026ivx}.}.  Slow-roll inflation arises
generically for kinetic-energy-dominated initial data at the bounce
\cite{Li:2019ipm}, and the background evolution exhibits universal features
for a broad class of inflationary potentials \cite{Saeed:2024xhk}.  This
asymmetric pre-bounce structure makes mLQC-I especially interesting for
studying possible quantum-gravitational imprints in primordial perturbations.

Cosmological perturbations in mLQC-I have consequently been investigated in
both the dressed-metric and hybrid approaches.  Within the dressed-metric
framework, the primordial power spectrum was studied in
Ref.~\cite{Agullo:2018wbf}, followed by a systematic comparison of the
scalar and tensor spectra of LQC, mLQC-I, and mLQC-II
\cite{Li:2019qzr}.  Within the hybrid framework, the perturbative
construction was extended to mLQC-I in Ref.~\cite{Gomar:2020orw}, while
the corresponding scalar and tensor effective masses and their relation to
the dressed-metric description were analyzed in
Ref.~\cite{Garcia-Quismondo:2020wna}.  The primordial scalar spectra of
the three loop cosmologies were compared within the hybrid framework in
Ref.~\cite{Li:2020mfi}.  The pre-bounce perturbation dynamics and the
selection of stable initial states were investigated further in
Ref.~\cite{Pan:2025ppa}.  More recently, high-accuracy mode functions,
ultraviolet regularity, and particle creation in the hybrid approach to
mLQC-I have been studied in Ref.~\cite{Saeed:2026vxw}.

Beyond the power spectrum, primordial non-Gaussianity provides an independent
and potentially more sensitive probe of pre-inflationary physics.  In
single-field slow-roll inflation the primordial bispectrum is strongly
constrained by the approximate de Sitter symmetries and is normally small
\cite{Maldacena:2002vr}.  Quantum-bounce dynamics can lead to qualitatively
different behavior.  Primordial non-Gaussianity in standard LQC and related
bouncing scenarios has been investigated in
Refs.~\cite{Agullo:2015aba,Agullo:2017eyh,Zhu:2017onp,Wu:2018sbr,
Agullo:2020cvg,Agullo:2020fbw,Agullo:2021oqk}.  These studies showed that
the bounce can generate strongly scale-dependent and oscillatory
non-Gaussianity, particularly on scales that are sensitive to the
pre-inflationary quantum-gravity regime.  To our knowledge, however, a
corresponding nonlinear bispectrum calculation has not previously been
carried out for either mLQC-I or mLQC-II.

The observational interpretation of such a primordial bispectrum requires
some care.  A large pointwise value of $f_{\rm NL}$ need not translate into
a large CMB bispectrum when the primordial signal oscillates on momentum
scales much shorter than those resolved by the radiation-transfer kernels.
For standard LQC, Roshna and Sreenath showed explicitly that rapid primordial
oscillations can be strongly averaged in the reduced CMB bispectrum
\cite{Roshna:2023}.  On the other hand, smoother bouncing bispectrum
templates can survive the transfer projection much more efficiently and may
be strongly constrained by CMB data \cite{vanTent:2023}.  It is therefore
important to determine not only the magnitude but also the momentum-space
phase and triangle dependence of the primordial bispectrum.

A logically distinct issue is whether the perturbative expansion itself
remains under control when the primordial three-point function becomes
large.  Agullo, Bolliet and Sreenath (ABS) addressed this question in standard 
LQC by computing the leading nonlinear correction to the curvature
power spectrum and comparing it with the leading-order result
\cite{Agullo:2017eyh}.  Their analysis demonstrated that a large primordial
bispectrum does not by itself imply a breakdown of the correlation-function
expansion, because the correction to the two-point function contains
additional powers of the small scalar power spectrum.  This motivates the
diagnostic
\begin{equation}
 \Xi(k)\equiv
 \left|
 \frac{\Delta{\cal P}_{\cal R}(k)}
      {{\cal P}_{\cal R}^{(0)}(k)}
 \right|,
 \label{eq:Xi-def}
\end{equation}
%\begingroup\color{blue}\label{rv3}
with $\Xi\ll1$ testing the smallness of the retained nonlinear-conversion
correction on the stated domain.
%\par\endgroup

The purpose of the present work is to extend both aspects of this nonlinear
analysis to mLQC-I within the hybrid approach.  First, we construct the cubic
interaction Hamiltonian in the same hybrid canonical variables used in the
quadratic theory and derive the exact finite-momentum tree-level bispectrum.
The ordering of operations is important: the classical quadratic and cubic
Hamiltonians are first written in a common canonical representation without
using the classical Friedmann constraint, the mLQC-I regularization is then
implemented, and only afterward is the effective background constraint used.
Second, we evaluate the ABS
correlation-function diagnostic for two physically distinct initial-state 
prescriptions: a Gaussian state specified
directly at the bounce and the de Sitter vacuum selected in the remote
contracting phase.  We also retain the complete second-order relation
between the spatially flat-gauge field perturbation and the comoving
curvature perturbation and construct its minimal mLQC-I effective
continuation, providing the basis for future nonlinear calculations in the
contracting phase and near the bounce \footnote{We emphasize that the correlation-function criterion employed here is more
limited than a general proof of the absence of strong coupling.  Genuine
$O\left(\left[H_I^{(3)}\right]^2\right)$ loop corrections, quartic and higher interactions,
composite-operator renormalization, unrestricted ultraviolet extensions,
and possible effective-field-theory cutoffs are logically distinct issues \cite{Saeed:2026xxx}.}.

Our results show that these two initial-state prescriptions lead to very
different degrees of cancellation sensitivity in the retained correlation-function
diagnostic. For the bounce-defined state, the finite-window absolute-integrand
estimate is below $6\times10^{-6}$ at the five sampled momenta.
%
%\label{v5removed:2}
%
For the remote
de Sitter state, the signed correction is more oscillatory and
cancellation dominated, decreasing strongly as the external scale approaches
the bounce-curvature scale.  At the same time, the primordial bispectrum can
be very large pointwise.  For the bounce-defined state the phase root-mean-square (RMS)
equilateral $f_{\rm NL}$ is typically of order $10^3$--$10^4$ over the
displayed range, while the remote prescription can give much larger values.

One of our central numerical results is that the rapidly varying part of the
bispectrum is approximately organized by the total momentum
$k_t\equiv k_1+k_2+k_3$, with a nearly universal oscillation period
$\Delta k_t\simeq5.7\times10^{-4}m_{\rm P}$ over the triangle configurations
studied, whereas the momentum-triangle envelope and relative phase vary much
more slowly.  This separation motivates a general numerical template for the
primordial non-Gaussianity.  For the illustrative bounce-to-present mapping
used in our projection analysis, the corresponding oscillations are much
faster than the CMB transfer scales, and our thin-shell and acoustic-transfer
diagnostics indicate substantial phase averaging.  
%These projection calculations are
%intended as diagnostics rather than as substitutes for a full Boltzmann-code analysis.

% We emphasize that the correlation-function criterion employed here is more
% limited than a general proof of the absence of strong coupling.  Genuine
% $O\left(\left[H_I^{(3)}\right]^2\right)$ loop corrections, quartic and higher interactions,
% composite-operator renormalization, unrestricted ultraviolet extensions,
% and possible effective-field-theory cutoffs are logically distinct issues.

The remainder of the paper is organized as follows.  In Sec.~II we write the
classical quadratic and cubic Hamiltonians in a common set of hybrid
canonical variables and present the complete nonlinear relation between the
field perturbation and the co-moving curvature perturbation ${\cal R}$. 
In Sec.~III we implement the mLQC-I
effective quantization, construct the cubic interaction Hamiltonian, derive
the minimal effective continuation of the nonlinear curvature perturbation,
and examine formal symmetry and the classical limit.  Sec.~IV derives the exact finite-momentum cubic vertex,
tree-level bispectrum, and the retained nonlinear-conversion
diagnostic.  In Sec.~V we specify the bounce-defined and remote-contracting
initial states and evaluate their perturbativity properties.  Sec.~VI
studies the primordial power spectrum and bispectrum, the rapid $k_t$
oscillation, the general momentum-triangle template, and CMB-projection
diagnostics.  Our conclusions are summarized in Sec.~VII. Four appendices provide useful algebraic relations, the canonical
completion, interpolation details, and independent numerical validation. 

Throughout this paper we set $\hbar=c=1$ and use $\kappa=8\pi G$.  We denote
the unreduced Planck mass and Planck time by
$m_{\rm P}\equiv G^{-1/2}$ and $t_{\rm P}\equiv G^{1/2}$, respectively.

\section{Classical Hamiltonians in the hybrid canonical variables}
\renewcommand{\theequation}{2.\arabic{equation}}
\setcounter{equation}{0}
\lb{sec:classical}

We write
\begin{equation}
 v=a^3,
 \quad
\Omega\equiv\Omega_{\rm can}=vb=\gamma a^3H_{\rm cl},
 \label{eq:Omega-def}
\end{equation}
and use the fiducial Euclidean convention
\begin{equation}
 \partial^2\equiv\delta^{ij}\partial_i\partial_j,
 \quad
 (\partial f)^2\equiv\delta^{ij}(\partial_i f)(\partial_j f),
 \label{eq:spatial-derivative-convention}
\end{equation}
for spatial derivatives in the transformed hybrid Hamiltonian.
Here $H_{\rm cl}=b/\gamma$ is the classical canonical relation; it is not
the effective mLQC-I rate. Throughout the classical phase-space substitutions,
$\Omega$ means $\Omega_{\rm can}$. The rate-based comparison variable
$\Omega_H=\gamma vH_{\rm eff}$ will be used explicitly only in
Sec.~\ref{subsec:GR-limit}.

\subsection{Quadratic Hamiltonian}
\lb{subsec:H2cl}

The quadratic Hamiltonian in the Mukhanov--Sasaki (MS) canonical pair
$(\nu,\pi_\nu)$ can be written as \cite{Fernandez-Mendez:2012poe,Fernandez-Mendez:2013jqa,Fernandez-Mendez:2014raa}
\bq
H_{\rm cl}^{(2)}
=\frac{N}{2a}\int d^3x\left[
\pi_\nu^2+(\partial\nu)^2+m_{\rm cl}^2\nu^2
\right].
\lb{eq:H2cl}
\eq
We use the off-shell classical mass in the canonical variables
$(a,\pi_a;\phi,p_\phi)$,
\bqn
m_{\rm cl}^2
&=&-\frac{27p_\phi^4}{2\pi_a^2a^6}
+\frac{5\kappa p_\phi^2}{2a^4}
+\frac{9p_\phi^2V}{\pi_a^2}
-\frac{12a p_\phi V_{,\phi}}{\pi_a}\nb\\
&&
-\frac{\kappa^2\pi_a^2}{72a^2}
+a^2V_{,\phi\phi}-\frac{\kappa}{2}a^2V.
\lb{eq:mcl-pia}
\eqn
No use of the classical background Hamiltonian constraint is made in
Eq.~(\ref{eq:mcl-pia}).  This is the form that will be carried into the
effective construction.

Using the classical relation
\bq
\pi_a=-\frac{6\Omega}{\kappa\gamma a},
\lb{eq:pia-Omega}
\eq
Eq.~(\ref{eq:mcl-pia}) can equivalently be written as
\bqn
m_{\rm cl}^{2}
&=&-\frac{3\kappa^2\gamma^2p_\phi^4}
{8v^{4/3}\Omega^2}
+\frac{5\kappa p_\phi^2}{2v^{4/3}}
+\frac{\kappa^2\gamma^2v^{2/3}p_\phi^2V}{4\Omega^2}\nb\\
&&
+\frac{2\kappa\gamma v^{2/3}p_\phi}{\Omega}V_{,\phi}
-\frac{\Omega^2}{2\gamma^2v^{4/3}}
+v^{2/3}V_{,\phi\phi}\nb\\
&&
-\frac{\kappa}{2}v^{2/3}V.
\lb{eq:mcl-Omega}
\eqn

\subsection{Classical cubic Hamiltonian in the spatially flat gauge}
\lb{subsec:H3-flat}

A convenient off-shell third-order Hamiltonian for scalar perturbations in
the spatially flat gauge is given in Ref.~\cite{Agullo:2017eyh}.  The
background Friedmann constraint is not used to simplify either the quadratic
or cubic coefficients at this stage, which is essential for applying the
mLQC-I regularization consistently.  We also retain explicitly the momentum
redefinition made before the displayed cubic Hamiltonian in that reference,
writing
\bq
\delta\bar p_\phi
=\delta p_\phi^{\rm old}
-\frac{3p_\phi^2}{a\pi_a}\delta\phi.
\lb{eq:pbar-def}
\eq
Thus, in the expression below, the momentum perturbation is understood as
$\delta\bar p_\phi$.

Before transforming to the hybrid variables, the cubic Hamiltonian reads
\begin{widetext}
\bqn
H_{\rm cl}^{(3)}=N\int d^3x\Bigg\{&&
\left(
\frac{9\kappa p_\phi^3}{4a^4\pi_a}
-\frac{27p_\phi^5}{2a^6\pi_a^3}
-\frac{3a^2p_\phi V_{,\phi\phi}}{2\pi_a}
+\frac{a^3V_{,\phi\phi\phi}}{6}
\right)(\delta\phi)^3
-\frac{3p_\phi}{2a^4\pi_a}(\delta\bar p_\phi)^2\delta\phi
-\frac{9p_\phi^3}{a^5\pi_a^2}\delta\bar p_\phi(\delta\phi)^2
\nb\\
&&-\frac{3a^2p_\phi}{2\pi_a}\delta\phi\,(\vec\partial\delta\phi)^2
+\frac{3p_\phi^2}{Na\pi_a}(\delta\phi)^2\partial^2\chi
+\frac{3a^2p_\phi}{2N^2\kappa\pi_a}\delta\phi(\partial^2\chi)^2
+\frac{3p_\phi^2}{Na\pi_a}\delta\phi\,\partial_i\chi\,\partial^i\delta\phi
\nb\\
&&+\frac{1}{N}\delta\bar p_\phi\,\partial_i\delta\phi\,\partial^i\chi
-\frac{3a^2p_\phi}{2N^2\kappa\pi_a}
\delta\phi\,\partial_i\partial_j\chi\,\partial^i\partial^j\chi
\Bigg\}.
\lb{eq:H3-Agullo}
\eqn
\end{widetext}
In Eq.~(\ref{eq:H3-Agullo}) the derivatives and index contractions are
written with the physical FLRW spatial metric.  They will be converted below
to the fiducial Euclidean convention used for the hybrid Hamiltonian.

The scalar shift potential $\chi$ follows from the first-order constraints.
After restoring the momentum variable in Eq.~(\ref{eq:pbar-def}), its Fourier
mode is
\bqn
\widetilde\chi_{\bf k}
&=&\frac{3N\kappa}{k^2a}
\Bigg[\left(
\frac{p_\phi}{2}
-\frac{a^5V_{,\phi}}{\kappa\pi_a}
-\frac{3p_\phi^3}{\kappa a^2\pi_a^2}
\right)\delta\widetilde\phi_{\bf k}
\nb\\
&& \hspace{1cm}
-\frac{p_\phi}{\kappa a\pi_a}
\delta\widetilde{\bar p}_{\phi{\bf k}}
\Bigg].
\lb{eq:chi-bar}
\eqn

\subsection{Canonical transformation to the hybrid pair}
\lb{subsec:canonical-transform}

The MS canonical pair used in the hybrid approach is related
to the variables above by
\bq
\nu=a\,\delta\phi,
\lb{eq:nu-transform}
\eq
and
\bq
\pi_\nu
=\frac{\delta\bar p_\phi}{a}
-\frac{\kappa\pi_a}{6a}\nu.
\lb{eq:pinu-transform}
\eq
Equivalently,
\bq
\delta\phi=\frac{\nu}{a},
 \quad
\delta\bar p_\phi
=a\left(\pi_\nu+\frac{\kappa\pi_a}{6a}\nu\right).
\lb{eq:inverse-transform}
\eq
Using Eq.~(\ref{eq:pia-Omega}), we define the convenient shorthand
\bq
{\cal P}_\nu
\equiv
\pi_\nu-\frac{\Omega}{\gamma v^{2/3}}\nu,
\lb{eq:Pnu-def}
\eq
so that
\bq
\delta\bar p_\phi=v^{1/3}{\cal P}_\nu.
\lb{eq:pbar-Pnu}
\eq
The canonical transformation on the full background-plus-perturbation
phase space requires a little care.  In particular, the momentum
redefinition in Eq.~(\ref{eq:pbar-def}) and the transformation in
Eqs.~(\ref{eq:nu-transform}) and (\ref{eq:pinu-transform}) should be
regarded as a single transformation from the original perturbative
pair $(\delta\phi,\delta p_\phi^{\rm old})$ to $(\nu,\pi_\nu)$.  Explicitly, we have
\begin{equation}
 \nu=a\,\delta\phi,
  \quad
 \pi_\nu
 =
 \frac{\delta p_\phi^{\rm old}}{a}
 -\frac{3p_\phi^2}{a^2\pi_a}\,\delta\phi
 -\frac{\kappa\pi_a}{6}\,\delta\phi,
 \label{eq:full-pert-canonical-map}
\end{equation}
or equivalently,
\begin{equation}
 \delta\phi=\frac{\nu}{a},
  \quad
 \delta p_\phi^{\rm old}
 =
 a\pi_\nu
 +
 \left(
 \frac{\kappa\pi_a}{6}
 +\frac{3p_\phi^2}{a^2\pi_a}
 \right)\nu .
 \label{eq:full-pert-canonical-map-inverse}
\end{equation}

Because the coefficients in this transformation depend on the homogeneous
variables, completing it canonically on the full phase space induces
quadratic perturbative redefinitions of the homogeneous variables.
As shown in Appendix~\ref{app:canonical-completion}, these shifts modify
the background Hamiltonian $H_{\rm cl}^{(0)}$ at second perturbative
order and therefore contribute to the quadratic hybrid Hamiltonian.
Their insertion into $H_{\rm cl}^{(2)}$, however, starts at fourth order,
whereas their insertion into $H_{\rm cl}^{(3)}$ starts at fifth order.
Furthermore, after gauge fixing and imposition of the first-order
constraints the reduced perturbative Hamiltonian starts at quadratic
order, $H_{\rm pert}=H^{(2)}+H^{(3)}+\cdots$, so there is no independent
first-order contribution that could combine with the quadratic
background shifts to generate an additional cubic term.  Consequently,
within the minimal extension in which $\nu=a\,\delta\phi$ remains linear,
the full canonical completion generates no additional contribution to
the cubic Hamiltonian in Eq.~(\ref{eq:H3cl-hybrid}), given below.

We also define
\bq
\vartheta\equiv\frac{\chi}{N}.
\lb{eq:vartheta-def}
\eq
After converting all spatial contractions to the convention of
Eq.~(\ref{eq:spatial-derivative-convention}), the classical cubic Hamiltonian
becomes
\begin{widetext}
\bqn
H_{\rm cl}^{(3)}
=N\int d^3x\Bigg\{&&
{\cal A}\nu^3
+{\cal B}\nu{\cal P}_\nu^2
-{\cal C}\nu^2{\cal P}_\nu
+{\cal B}\nu(\partial\nu)^2
-{\cal D}\left[
\nu^2\partial^2\vartheta
+\nu\,\partial_i\vartheta\,\partial^i\nu
\right]
+\frac{1}{v^{2/3}}{\cal P}_\nu\,
\partial_i\nu\,\partial^i\vartheta
\nb\\
&&+{\cal E}\nu\left[
(\partial_i\partial_j\vartheta)(\partial^i\partial^j\vartheta)
-(\partial^2\vartheta)^2
\right]
\Bigg\},
\lb{eq:H3cl-hybrid}
\eqn
\end{widetext}
where
\bqn
{\cal A}
&\equiv&-\frac{3\kappa^2\gamma p_\phi^3}{8v^2\Omega}
+\frac{\kappa^3\gamma^3p_\phi^5}{16v^2\Omega^3}
+\frac{\kappa\gamma p_\phi}{4\Omega}V_{,\phi\phi} +\frac{1}{6}V_{,\phi\phi\phi},
\nb\\
{\cal B}&\equiv& \frac{\kappa\gamma p_\phi}{4v^{2/3}\Omega},
 \quad
{\cal C}\equiv \frac{\kappa^2\gamma^2p_\phi^3}{4v^{4/3}\Omega^2},\nb\\
{\cal D}&\equiv&\frac{\kappa\gamma p_\phi^2}{2v^{4/3}\Omega},
 \quad
{\cal E}\equiv\frac{\gamma p_\phi}{4v^{2/3}\Omega}.
\lb{eq:DE-cl}
\eqn
The corresponding Fourier-space shift variable is
\bqn
\vartheta_{\bf k}
&=&\frac{3\kappa}{k^2}\Bigg\{
v^{-2/3}\left[
\frac{p_\phi}{2}
+\frac{\gamma v^2}{6\Omega}V_{,\phi}
-\frac{\kappa\gamma^2p_\phi^3}{12\Omega^2}
\right]\nu_{\bf k}\nb\\
&& ~~~~~~~~~~
+\frac{\gamma p_\phi}{6\Omega}
{\cal P}_{\nu{\bf k}}
\Bigg\}.
\lb{eq:vartheta-cl}
\eqn
Equations~(\ref{eq:H3cl-hybrid})--(\ref{eq:vartheta-cl}) are the classical
cubic Hamiltonian written entirely in the same canonical representation used
for the hybrid quadratic theory, before any loop regularization is applied.

\subsection{Nonlinear relation to the comoving curvature perturbation}
\label{subsec:classical-R}

The nonlinear relation between the spatially flat-gauge field perturbation
and the comoving curvature perturbation is a classical gauge transformation.
Because the same relation will be needed in a future analysis of the
contracting phase and of the bounce itself, we keep the complete
second-order expression rather than introducing only its late-time
super-Hubble reduction.  We first define
\begin{equation}
 z\equiv-\frac{6}{\kappa}\frac{p_\phi}{\pi_a}
 =\frac{\gamma a p_\phi}{\Omega}
 =\frac{a\dot\phi}{H}.
 \label{eq:z-classical}
\end{equation}
Since $\nu=a\,\delta\phi$, the linear relation is
\begin{equation}
 {\cal R}^{(1)}=-\frac{\nu}{z}.
 \label{eq:R-linear-classical}
\end{equation}
It is also useful to define the coefficient of the local quadratic term,
\begin{equation}
 C_{\cal R}
 =-\frac{3}{2}
 +\frac{3V_{,\phi}a^5}{\kappa p_\phi\pi_a}
 +\frac{\kappa}{4}\frac{z^2}{a^2}.
 \label{eq:CR-classical-canonical}
\end{equation}
Then the complete second-order transformation of
Refs.~\cite{Maldacena:2002vr,Agullo:2017eyh} is
\begin{widetext}
\begin{align}
 {\cal R}(\mathbf{x},t)
 ={}&-\frac{a}{z}\,\delta\phi
 +\left[
 -\frac{3}{2}
 +\frac{3V_{,\phi}a^5}{\kappa p_\phi\pi_a}
 +\frac{\kappa}{4}\frac{z^2}{a^2}
 \right]
 \left(\frac{a}{z}\delta\phi\right)^2
 -\frac{3a^2}{\kappa\pi_a}
 \frac{d}{dt}
 \left(\frac{a}{z}\delta\phi\right)^2-\frac{9a^4}{\kappa^2\pi_a^2}\frac{a^2}{z^2}
 (\partial\delta\phi)^2
 \nonumber\\
 &+\frac{9a^4}{\kappa^2\pi_a^2}\frac{a^2}{z^2}
 \partial^{-2}\partial_i\partial_j
 \left(\partial^i\delta\phi\,\partial^j\delta\phi\right)
 +\frac{3a^4}{\kappa\pi_a}\frac{a}{z}
 \partial_i\chi\,\partial^i\delta\phi
 -\frac{3a^4}{\kappa\pi_a}\frac{a}{z}
 \partial^{-2}\partial_i\partial_j
 \left(\partial^i\chi\,\partial^j\delta\phi\right).
 \label{eq:R-classical-full}
\end{align}
\end{widetext}
Here $\partial^{-2}$ denotes the inverse spatial Laplacian and
$\chi=N\vartheta$, with $\vartheta$ given in
Eq.~(\ref{eq:vartheta-cl}).  In Eq.~(\ref{eq:R-classical-full}) we retain
the convention of Ref.~\cite{Agullo:2017eyh}: spatial indices in the
second-order gauge transformation are raised and lowered with the physical
FLRW spatial metric $q^{(0)}_{ij}=a^2\delta_{ij}$.  This convention should
not be confused with the fiducial-Euclidean derivative convention introduced
in Eq.~(\ref{eq:spatial-derivative-convention}) for the transformed hybrid
Hamiltonian.  Equation~(\ref{eq:R-classical-full}), rather than its local
reduction below, is the relation relevant when the non-Gaussianity is
evaluated in the contracting phase or through the bounce: the
time-derivative, gradient, inverse-Laplacian, and shift terms cannot then be
discarded a priori.

\section{mLQC-I effective quantization}
\renewcommand{\theequation}{3.\arabic{equation}}
\setcounter{equation}{0}
\lb{sec:effective}

\subsection{Background effective Hamiltonian}
\lb{subsec:H0eff}

We start from the classical homogeneous Hamiltonian
\bq
H_{\rm cl}^{(0)}
=-\frac{3\Omega^2}{\kappa v\gamma^2}
+\frac{p_\phi^2}{2v}+vV,
\lb{eq:H0cl}
\eq
and the mLQC-I effective replacement
\bqn
\Omega^2\rightarrow\Omega_{\rm I}^2
&=&-\frac{v^2\gamma^2}{\lambda^2}
\left\{
\sin^2(\lambda b)
-\frac{\gamma^2+1}{4\gamma^2}\sin^2(2\lambda b)
\right\}. ~~~~~~~
\lb{eq:OmegaI}
\eqn
Thus, we find
\bq
H_{\rm eff}^{(0)}
=-\frac{3\Omega_{\rm I}^2}{\kappa v\gamma^2}
+\frac{p_\phi^2}{2v}+vV.
\lb{eq:H0eff}
\eq
It is useful to define
\bq
\Lambda_{\rm I}\equiv\frac{v}{2\lambda}\sin(2\lambda b).
\lb{eq:LambdaI}
\eq
The perturbative prescription used in the hybrid mLQC-I construction is
\bq
\frac{1}{\Omega^2}\longrightarrow\frac{1}{\Omega_{\rm I}^2},
 \quad
\frac{1}{\Omega}\longrightarrow
\frac{\Lambda_{\rm I}}{\Omega_{\rm I}^2}.
\lb{eq:hybrid-prescription}
\eq
The first rule is tied to the mLQC-I background regularization, while the
second is chosen so that the perturbative operator respects the background
superselection structure \cite{Li:2020mfi}.  The same quadratic prescription
is used in our particle-creation analysis \cite{Saeed:2026vxw}.

\subsection{Quadratic effective Hamiltonian and consistency check}
\lb{subsec:H2eff}

Applying Eqs.~(\ref{eq:OmegaI}) and (\ref{eq:hybrid-prescription}) directly
to the off-shell expression (\ref{eq:mcl-Omega}) gives
\bqn
m_{\rm eff,raw}^{2}
&=&-\frac{3\kappa^2\gamma^2p_\phi^4}
{8v^{4/3}\Omega_{\rm I}^2}
+\frac{5\kappa p_\phi^2}{2v^{4/3}}\nb\\
&&+\frac{\kappa^2\gamma^2v^{2/3}p_\phi^2V}
{4\Omega_{\rm I}^2}
+\frac{2\kappa\gamma v^{2/3}p_\phi\Lambda_{\rm I}}
{\Omega_{\rm I}^2}V_{,\phi}\nb\\
&&-\frac{\Omega_{\rm I}^2}{2\gamma^2v^{4/3}}
+v^{2/3}V_{,\phi\phi}
-\frac{\kappa}{2}v^{2/3}V.
\lb{eq:meff-raw}
\eqn
The effective mass used in the hybrid mLQC-I mode analysis \cite{Li:2020mfi,Saeed:2026vxw} is
\bqn
m_{\rm eff}^{2}
&=&\frac{\kappa p_\phi^2}{6v^{4/3}}
\left(19-\frac{3\kappa\gamma^2p_\phi^2}{\Omega_{\rm I}^2}\right)
\nb\\
&&+v^{2/3}\left[
V_{,\phi\phi}
+\frac{2\kappa\gamma p_\phi\Lambda_{\rm I}}
{\Omega_{\rm I}^2}V_{,\phi}
-\frac{2\kappa}{3}V
\right].
\lb{eq:meff-hybrid}
\eqn
The two expressions differ only by the effective background constraint:
\bq
m_{\rm eff,raw}^{2}-m_{\rm eff}^{2}
=
\frac{\kappa\left(2\Omega_{\rm I}^2
+3\kappa\gamma^2p_\phi^2\right)}
{12v^{1/3}\Omega_{\rm I}^2}
H_{\rm eff}^{(0)}.
\lb{eq:quadratic-check}
\eq
Consequently, once the effective background constraint
$H_{\rm eff}^{(0)}=0$ is imposed \emph{after} quantization, the off-shell
canonical construction reproduces precisely the hybrid effective mass.  This
provides an important consistency check on the ordering of the procedure.
Thus, the quadratic effective Hamiltonian reads
\bq
H_{\rm eff}^{(2)}
=\frac{N}{2v^{1/3}}\int d^3x\left[
\pi_\nu^2+(\partial\nu)^2+m_{\rm eff}^2\nu^2
\right].
\lb{eq:H2eff}
\eq
For conformal time, $N=a=v^{1/3}$, Hamilton's equations give
\begin{equation}
 \nu'=\pi_\nu,
 \quad
 \pi_\nu'=\partial^2\nu-m_{\rm eff}^2\nu.
 \label{eq:Hamilton-linear}
\end{equation}
Thus, each Fourier mode satisfies
\begin{equation}
 \nu_k''+\left(k^2+m_{\rm eff}^2\right)\nu_k=0.
 \label{eq:mode-equation}
\end{equation}
Here and below, the subscript $k$ labels the Fourier mode of the corresponding
quantity.
%; it does not denote differentiation with respect to $k$.

\subsection{Cubic extension of the hybrid prescription}
\lb{subsec:H3-quantization}

At cubic order one encounters odd inverse powers of the background
momentum that are not explicitly required in the quadratic theory.
Rather than introducing an independent polymerization prescription for
each such power, we adopt the same factor-ordering and
superselection-sector prescription used in the hybrid quantization of
the mLQC-I background and its perturbations
\cite{Garcia-Quismondo:2019dwa,Li:2020mfi}.

For convenience, we define the background operators
\bq
\lb{eq3.10}
\widehat X_{\rm I}\equiv\widehat{\Omega_{\rm I}^{2}},
 \quad
\widehat L_{\rm I}\equiv\widehat{\Lambda_{\rm I}},
\eq
where $\widehat X_{\rm I}$ is the operator representing the squared
background momentum and $\widehat L_{\rm I}$ is the corresponding odd
operator chosen so as to preserve the same step-four superselection
sectors; the underlying background-operator construction and its
superselection structure are discussed in
Refs.~\cite{Garcia-Quismondo:2019dwa,Li:2020mfi}.

Guided by these constructions, we extend the same ordering principle to
arbitrary odd integer powers.  Even integer powers are represented by
powers of $\widehat X_{\rm I}$, whereas odd integer powers are represented
symmetrically as
\bq
\lb{eq3.11}
\widehat{\Omega^{\,2l+1}}
=
\left|\widehat X_{\rm I}\right|^{l/2}
\widehat L_{\rm I}
\left|\widehat X_{\rm I}\right|^{l/2},
 \quad
l\in{\mathbb Z}.
\eq
% \begingroup\color{blue}\label{rv10}
% Spectral powers require a chosen self-adjoint realization of
% $\widehat X_{\rm I}$ on the selected background sector. We assume such a
% realization and restrict inverse powers to its kernel-orthogonal spectral
% domain, requiring the relevant inverse moments to exist. A common dense
% invariant domain for $\widehat L_{\rm I}$ and all ordered products is an
% additional assumption. 
Thus, for $l=-1$, we take
\bq
\lb{eq3.12}
\widehat{\Omega^{-1}}
=
\left|\widehat X_{\rm I}\right|^{-1/2}
\widehat L_{\rm I}
\left|\widehat X_{\rm I}\right|^{-1/2}.
\eq
For a commuting sharply peaked symbol the odd inverse is
$\Lambda_{\rm I}/|X_{\rm I}|$, where $X_{\rm I}=\Omega_{\rm I}^2$.
On the positive-density constraint surface,
$X_{\rm I}=\kappa\gamma^2v^2\rho/3>0$, it reduces to
%\par\endgroup
\bq
\lb{eq3.13}
\frac{1}{\Omega}
\longrightarrow
\frac{\Lambda_{\rm I}}{\Omega_{\rm I}^{2}},
\eq
which reproduces the prescription already used in the hybrid mLQC-I
construction and in the particle-creation analysis of Ref.~\cite{Saeed:2026vxw}.

The cubic Hamiltonian additionally contains the inverse odd power
$\Omega^{-3}$.  Taking $l=-2$ in the same operator prescription gives
\bq
\lb{eq3.14}
\widehat{\Omega^{-3}}
=
\left|\widehat X_{\rm I}\right|^{-1}
\widehat L_{\rm I}
\left|\widehat X_{\rm I}\right|^{-1}.
\eq
Consequently, for sharply peaked semiclassical background states, we find
\begin{equation}
 {
 \frac{1}{\Omega^3}
 \longrightarrow
 \frac{\Lambda_{\rm I}}{\Omega_{\rm I}^{4}},
 }
 \lb{eq:Omega3-prescription}
\end{equation}
which fixes the inverse-cubic term appearing in the coefficient
${\cal A}$.

The same prescription also fixes the positive odd power of $\Omega$
that occurs in the hybrid momentum combination.  Indeed, setting
$l=0$ we find
\bq
\lb{eq3.16}
\widehat{\Omega}=\widehat L_{\rm I},
\eq
and therefore, at the effective level, we obtain
\begin{equation}
 {
 {\cal P}_{\nu{\rm I}}
 \equiv
 \pi_\nu
 -
 \frac{\Lambda_{\rm I}}{\gamma v^{2/3}}\nu .
 }
 \lb{eq:PnuI-def}
\end{equation}
We shall use the shorthand
\begin{equation}
 \Gamma_{\rm I}
 \equiv
 \frac{\Lambda_{\rm I}}{\gamma v^{2/3}},
 \label{eq:GammaI}
\end{equation}
so that ${\cal P}_{\nu{\rm I}}=\pi_\nu-\Gamma_{\rm I}\nu$.
Thus, replacements
$\Omega^{-1}\rightarrow\Lambda_{\rm I}/\Omega_{\rm I}^{2}$,
$\Omega^{-3}\rightarrow\Lambda_{\rm I}/\Omega_{\rm I}^{4}$, and
$\Omega\rightarrow\Lambda_{\rm I}$ are not treated as independent
polymerization choices.  They follow from the same odd-power
factor-ordering prescription.

A useful organizational rule is to retain the composite shift variable
$\vartheta_{\rm I}$ in factorized form.  In particular, products
generated by $\vartheta_{\rm I}^{\,2}$ should not be expanded and then
reinterpreted as independent polymerizations of higher inverse powers.
The factors entering $\vartheta_{\rm I}$ are quantized first according
to the preceding prescription, and their products are subsequently
formed.  This preserves the operator factorization inherited from the
classical expression.

We stress that this construction does not establish a unique
quantization of the cubic theory.  Rather, it is the extension selected
by maintaining the same factor ordering and superselection-sector
prescription used in the hybrid mLQC-I quantization underlying the
quadratic theory \cite{Garcia-Quismondo:2019dwa,Li:2020mfi}.

 \subsection{Effective cubic Hamiltonian}
 \lb{subsec:H3eff}

 With the preceding prescription, we define
 \bqn
 {\cal A}_{\rm I}
 &=&-\frac{3\kappa^2\gamma p_\phi^3}{8v^2}
 \frac{\Lambda_{\rm I}}{\Omega_{\rm I}^2}
 +\frac{\kappa^3\gamma^3p_\phi^5}{16v^2}
 \frac{\Lambda_{\rm I}}{\Omega_{\rm I}^4}
 \nb\\
 &&+\frac{\kappa\gamma p_\phi}{4}
 \frac{\Lambda_{\rm I}}{\Omega_{\rm I}^2}V_{,\phi\phi}
 +\frac{1}{6}V_{,\phi\phi\phi},\nb\\
 {\cal B}_{\rm I}
 &=&\frac{\kappa\gamma p_\phi}{4v^{2/3}}
 \frac{\Lambda_{\rm I}}{\Omega_{\rm I}^2},
  \quad
 {\cal C}_{\rm I}
 =\frac{\kappa^2\gamma^2p_\phi^3}
 {4v^{4/3}\Omega_{\rm I}^2},\nb\\
 {\cal D}_{\rm I}
 &=&\frac{\kappa\gamma p_\phi^2}{2v^{4/3}}
 \frac{\Lambda_{\rm I}}{\Omega_{\rm I}^2},
  \quad
 {\cal E}_{\rm I}
 =\frac{\gamma p_\phi}{4v^{2/3}}
 \frac{\Lambda_{\rm I}}{\Omega_{\rm I}^2}.
 \lb{eq:DEI}
 \eqn
 Then, the effective shift variable is
 \begin{widetext}
 \bqn
 \vartheta_{{\rm I},{\bf k}}
 =\frac{3\kappa}{k^2}\Bigg\{&&
 \left[
 \frac{p_\phi}{2v^{2/3}}
 +\frac{\gamma v^{4/3}}{6}
 \frac{\Lambda_{\rm I}}{\Omega_{\rm I}^2}V_{,\phi}
 -\frac{\kappa\gamma^2p_\phi^3}
 {12v^{2/3}\Omega_{\rm I}^2}
 \right]\nu_{\bf k}
 +\frac{\gamma p_\phi}{6}
 \frac{\Lambda_{\rm I}}{\Omega_{\rm I}^2}
 {\cal P}_{\nu{\rm I},{\bf k}}
 \Bigg\}.
 \lb{eq:vartheta-I}
 \eqn
The resulting effective cubic Hamiltonian takes the compact form
 \bqn
 H_{\rm eff}^{(3)}
 =N\int d^3x\Bigg\{&&
 {\cal A}_{\rm I}\nu^3
 +{\cal B}_{\rm I}\nu{\cal P}_{\nu{\rm I}}^2
 -{\cal C}_{\rm I}\nu^2{\cal P}_{\nu{\rm I}}
 +{\cal B}_{\rm I}\nu(\partial\nu)^2
 -{\cal D}_{\rm I}\left[
 \nu^2\partial^2\vartheta_{\rm I}
 +\nu\,\partial_i\vartheta_{\rm I}\,\partial^i\nu
 \right]
 +\frac{1}{v^{2/3}}{\cal P}_{\nu{\rm I}}
 \partial_i\nu\,\partial^i\vartheta_{\rm I}
 \nb\\
 &&+{\cal E}_{\rm I}\nu\left[
 (\partial_i\partial_j\vartheta_{\rm I})
 (\partial^i\partial^j\vartheta_{\rm I})
 -(\partial^2\vartheta_{\rm I})^2
 \right]
 \Bigg\}.
 \lb{eq:H3eff}
 \eqn
 \end{widetext}
 The full effective Hamiltonian through cubic order is therefore
 \bq
 H_{\rm eff}
 =H_{\rm eff}^{(0)}+H_{\rm eff}^{(2)}+H_{\rm eff}^{(3)}+\cdots.
 \lb{eq:Heff-total}
 \eq
 The classical limit is immediate:
 \bq
 \Omega_{\rm I}^2\rightarrow\Omega^2,
  \quad
 \Lambda_{\rm I}\rightarrow\Omega,
  \quad
 H_{\rm eff}^{(3)}\rightarrow H_{\rm cl}^{(3)}.
 \lb{eq:classical-limit}
 \eq

\subsection{Effective continuation of the nonlinear curvature perturbation}
\label{subsec:R-eff}

The complete relation in Eq.~(\ref{eq:R-classical-full}) is a classical
second-order gauge transformation.  By itself it does not define a unique
quantum-corrected curvature perturbation in the Planck regime.  Nevertheless,
for the purpose of extending the same hybrid prescription consistently to
all terms that will enter a future pre-bounce and near-bounce
non-Gaussianity calculation, we define here a \emph{minimal term-by-term
effective continuation}.  The factorization specified below is part of this prescription: in general,
effective continuation does not preserve products or inverses. Its derivation
from the full second-order effective constraint algebra, identification with
a geometrically defined comoving curvature perturbation, and regularity must
be established separately. The expressions below apply only where their
denominators are nonzero. No new polymerization rule is introduced.

From
\begin{equation}
 z=\frac{\gamma v^{1/3}p_\phi}{\Omega},
\end{equation}
the replacement
$\Omega^{-1}\rightarrow\Lambda_{\rm I}/\Omega_{\rm I}^2$ gives
\begin{equation}
 z_{\rm I}
 \equiv
 \gamma v^{1/3}p_\phi
 \frac{\Lambda_{\rm I}}{\Omega_{\rm I}^2}.
 \label{eq:zI}
\end{equation}
The factors of $\pi_a^{-1}$ and $\pi_a^{-2}$ appearing in the full
second-order transformation can be written classically as
\begin{equation}
 \frac{1}{\pi_a}
 =-\frac{\kappa\gamma a}{6}\frac{1}{\Omega},
 \quad
 \frac{1}{\pi_a^2}
 =\frac{\kappa^2\gamma^2a^2}{36}\frac{1}{\Omega^2}.
\end{equation}
Their effective continuations are therefore
\bqn %egin{align}
 \Pi_{1{\rm I}}
 &\equiv&
 \left(\frac{1}{\pi_a}\right)_{\rm eff}
 =-\frac{\kappa\gamma a}{6}
 \frac{\Lambda_{\rm I}}{\Omega_{\rm I}^2},\nb\\
 \Pi_{2{\rm I}}
 &\equiv&
 \left(\frac{1}{\pi_a^2}\right)_{\rm eff}
 =\frac{\kappa^2\gamma^2a^2}{36}
 \frac{1}{\Omega_{\rm I}^2}.
 \label{eq:Pi2I}
\eqn %nd{align}
Notice that $\Pi_{2{\rm I}}\neq(\Pi_{1{\rm I}})^2$ in general.  This is
intentional: the even inverse power $\Omega^{-2}$ is represented by the
inverse of $\widehat X_{\rm I}$, whereas $\Omega^{-1}$ is represented by
the sector-preserving odd operator in Eq.~(\ref{eq3.12}).  Thus
Eq.~(\ref{eq:Pi2I}) follows the same even/odd
factor-ordering prescription already used in the Hamiltonian.
For the shift contribution we use the already-defined effective variable
\begin{equation}
 \chi_{\rm I}\equiv N\vartheta_{\rm I},
 \label{eq:chiI}
\end{equation}
where $\vartheta_{{\rm I},{\bf k}}$ is given in
Eq.~(\ref{eq:vartheta-I}).  Finally, direct continuation of the local
coefficient gives
\begin{equation}
 C_{{\cal R},{\rm I}}
 =-\frac{3}{2}
 -\frac{\gamma v^2V_{,\phi}\Lambda_{\rm I}}
 {2p_\phi\Omega_{\rm I}^2}
 +\frac{\kappa\gamma^2p_\phi^2\Lambda_{\rm I}^2}
 {4\Omega_{\rm I}^4}.
 \label{eq:CRI}
\end{equation}

With these definitions, the minimal effective continuation of the
\emph{complete} second-order curvature perturbation is
\begin{widetext}
\begin{align}
 {\cal R}_{\rm eff}(\mathbf{x},t)
 ={}&-\frac{a}{z_{\rm I}}\delta\phi
 +C_{{\cal R},{\rm I}}
 \left(\frac{a}{z_{\rm I}}\delta\phi\right)^2
 -\frac{3a^2}{\kappa}\Pi_{1{\rm I}}
 \frac{d}{dt}
 \left(\frac{a}{z_{\rm I}}\delta\phi\right)^2
 -\frac{9a^4}{\kappa^2}\Pi_{2{\rm I}}
 \frac{a^2}{z_{\rm I}^2}
 (\partial\delta\phi)^2
 +\frac{9a^4}{\kappa^2}\Pi_{2{\rm I}}
 \frac{a^2}{z_{\rm I}^2}
 \partial^{-2}\partial_i\partial_j
 \left(\partial^i\delta\phi\,\partial^j\delta\phi\right)
 \nonumber\\
 &+\frac{3a^4}{\kappa}\Pi_{1{\rm I}}\frac{a}{z_{\rm I}}
 \partial_i\chi_{\rm I}\,\partial^i\delta\phi
 -\frac{3a^4}{\kappa}\Pi_{1{\rm I}}\frac{a}{z_{\rm I}}
 \partial^{-2}\partial_i\partial_j
 \left(\partial^i\chi_{\rm I}\,\partial^j\delta\phi\right).
 \label{eq:R-eff-full}
\end{align}
\end{widetext}
Equation~(\ref{eq:R-eff-full}) retains precisely the structures that become
important away from the late-time super-Hubble regime: the local quadratic
piece, the time-derivative term, the two spatial-gradient terms, and the two
shift-dependent terms.  The spatial derivatives in this equation follow the
same physical-FLRW convention as in Eq.~(\ref{eq:R-classical-full}); the
scalar $\chi_{\rm I}=N\vartheta_{\rm I}$ itself is independent of this
notational choice.  Equation~(\ref{eq:R-eff-full}) is therefore the natural
starting point for testing the nonlinear correlation functions in the
contracting phase and through the bounce.

For comparison with the late-time calculation performed in the present
paper, define
\begin{equation}
 q_{\rm I}\equiv-\frac{\nu}{z_{\rm I}}.
 \label{eq:qI}
\end{equation}

On the expanding classical branch, with all relevant modes super-Hubble
and derivative/shift terms negligible relative to the retained local term,
Eq.~(\ref{eq:R-eff-full}) is approximated by

\begin{equation}
 {\cal R}_{\rm eff}
 =q_{\rm I}+C_{{\cal R},{\rm I}}q_{\rm I}^2+\cdots.
 \label{eq:R-eff-local}
\end{equation}
In the low-curvature limit,
$\Omega_{\rm I}^2\rightarrow\Omega^2$ and
$\Lambda_{\rm I}\rightarrow\Omega$, so that
$z_{\rm I}\rightarrow z$,
$\Pi_{1{\rm I}}\rightarrow\pi_a^{-1}$,
$\Pi_{2{\rm I}}\rightarrow\pi_a^{-2}$,
$\chi_{\rm I}\rightarrow\chi$, and
Eq.~(\ref{eq:R-eff-full}) reduces to the complete classical relation
(\ref{eq:R-classical-full}).

We emphasize that Eq.~(\ref{eq:R-eff-full}) is a minimal effective
continuation selected by the same hybrid rules used in the quadratic and
cubic Hamiltonians; it is not a derivation of a unique nonlinear quantum gauge
transformation.  Establishing such uniqueness would require the full
second-order effective constraint and gauge-transformation algebra.
%\label{v5removed:5}
  The
present late-time results do not depend on this stronger construction,
whereas the full expression above makes explicit which terms must be retained
in a future pre-bounce analysis.

\subsection{Formal symmetry} % and fiducial-cell assumptions}
\lb{subsec:hermiticity-fiducial}

Before using $H_{\rm eff}^{(3)}$ as the interaction Hamiltonian, it is
useful to state the assumptions under which the operator prescription is
formally symmetric and to distinguish density conventions from dependence on
the infrared regulator.

We first consider the background operators.  In
Sec.~\ref{subsec:H3-quantization} we introduced
\begin{equation}
 \widehat X_{\rm I}\equiv\widehat{\Omega_{\rm I}^{2}},
 \quad
 \widehat L_{\rm I}\equiv\widehat{\Lambda_{\rm I}},
 \label{eq:XI-LI-hermiticity}
\end{equation}
and represented odd inverse powers by symmetric sandwich orderings.
In particular,
\begin{equation}
 \widehat{\Omega^{-1}}
 =
 \left|\widehat X_{\rm I}\right|^{-1/2}
 \widehat L_{\rm I}
 \left|\widehat X_{\rm I}\right|^{-1/2},
 \label{eq:Omega-minus1-herm}
\end{equation}
and
\begin{equation}
 \widehat{\Omega^{-3}}
 =
 \left|\widehat X_{\rm I}\right|^{-1}
 \widehat L_{\rm I}
 \left|\widehat X_{\rm I}\right|^{-1}.
 \label{eq:Omega-minus3-herm}
\end{equation}
%\begingroup\color{blue}\label{rv12}
The sandwiches are formally symmetric on the assumed common invariant
product domain. Their spectral factors use the self-adjoint background
realization assumed in Sec.~\ref{subsec:H3-quantization}, with the inverse
kernel excluded.  The displayed effective inverse symbols are used
on the positive-density constraint surface, not across the kernel or on
arbitrary off-shell points.
%\par\endgroup

Products involving powers of the volume and inverse-background
operators are ordered symmetrically.  For example, a classical product
$v^s F$ is represented schematically by
\begin{equation}
 v^s F
 \quad\longrightarrow\quad
 \widehat v^{\,s/2}\,
 \widehat F\,
 \widehat v^{\,s/2},
 \label{eq:background-sandwich}
\end{equation}
whenever $\widehat F$ does not commute with $\widehat v$.  The same
principle is applied to the background coefficients
${\cal A}_{\rm I},\ldots,{\cal E}_{\rm I}$.

There is an additional ordering issue in the matter sector whenever
$p_\phi$ multiplies a nontrivial function of $\phi$.  Such products are
taken in symmetric form.  In particular,
\begin{equation}
 p_\phi V_{,\phi\phi}
 \quad\longrightarrow\quad
 \frac{1}{2}
 \left\{
 \widehat p_\phi,
 V_{,\phi\phi}(\widehat\phi)
 \right\}.
 \label{eq:matter-Weyl-ordering}
\end{equation}
The remaining factors of $p_\phi$ commute with purely geometric
operators and therefore do not introduce an additional ordering
ambiguity.

The odd background momentum entering the perturbative
momentum combination is represented by
$\widehat L_{\rm I}$.  A symmetric realization of the corresponding
background coefficient is therefore
\begin{equation}
 \widehat\Gamma_{\rm I}
 \equiv
 \frac{1}{\gamma}\,
 \widehat v^{-1/3}
 \widehat L_{\rm I}
 \widehat v^{-1/3},
 \label{eq:GammaI-operator}
\end{equation}
whose sharply peaked effective value is
\begin{equation}
 \Gamma_{\rm I}
 =
 \frac{\Lambda_{\rm I}}
 {\gamma v^{2/3}}.
 \label{eq:GammaI-effective}
\end{equation}
Consequently, we find that
\begin{equation}
 \widehat{\cal P}_{\nu{\rm I}}
 =
 \widehat\pi_\nu
 -
 \widehat\Gamma_{\rm I}\widehat\nu
 \label{eq:PnuI-operator-herm}
\end{equation}
is symmetric, since background operators commute with the perturbative
canonical pair.

At cubic order an additional ordering choice is required because
$\widehat\nu$ and $\widehat{\cal P}_{\nu{\rm I}}$ do not commute.  We
adopt Weyl ordering for the perturbative monomials.  For three
operators $\widehat A$, $\widehat B$, and $\widehat C$, let
$(\widehat A_1,\widehat A_2,\widehat A_3)
=(\widehat A,\widehat B,\widehat C)$ and define
\begin{equation}
 {\cal W}[\widehat A\widehat B\widehat C]
 \equiv
 \frac{1}{3!}
 \sum_{\sigma\in S_3}
 \widehat A_{\sigma(1)}
 \widehat A_{\sigma(2)}
 \widehat A_{\sigma(3)} .
 \label{eq:Weyl-cubic}
\end{equation}
Thus, for example,
\begin{equation}
 \nu{\cal P}_{\nu{\rm I}}^2
 \quad\longrightarrow\quad
 {\cal W}
 \left[
 \widehat\nu
 \widehat{\cal P}_{\nu{\rm I}}
 \widehat{\cal P}_{\nu{\rm I}}
 \right],
 \label{eq:nuP2-ordering}
\end{equation}
and
\begin{equation}
 \nu^2{\cal P}_{\nu{\rm I}}
 \quad\longrightarrow\quad
 {\cal W}
 \left[
 \widehat\nu
 \widehat\nu
 \widehat{\cal P}_{\nu{\rm I}}
 \right].
 \label{eq:nu2P-ordering}
\end{equation}
The same prescription is applied to the cubic terms containing the
shift variable $\vartheta_{\rm I}$.

In Fourier space the latter may be written schematically as
\begin{equation}
 \widehat\vartheta_{{\rm I},{\bf k}}
 =
 \frac{3\kappa}{k^2}
 \left[
 \widehat F_{\nu{\rm I}}\,
 \widehat\nu_{\bf k}
 +
 \frac12\left\{\widehat F_{P{\rm I}},
 \widehat{\cal P}_{\nu{\rm I},{\bf k}}\right\}
 \right],
  \quad k\neq0,
 \label{eq:vartheta-operator-herm}
\end{equation}
where the background coefficients
$\widehat F_{\nu{\rm I}}$ and $\widehat F_{P{\rm I}}$ are themselves
defined with the symmetric background ordering described above. The
anticommutator symmetrizes the product because $\widehat F_{P{\rm I}}$
need not commute with $\widehat\Gamma_{\rm I}$. Products of the
remaining background-dependent factors are likewise symmetrized.
Since
\begin{equation}
 \widehat\nu_{\bf k}^{\dagger}
 =
 \widehat\nu_{-{\bf k}},
  \quad
 \widehat{\cal P}_{\nu{\rm I},{\bf k}}^{\dagger}
 =
 \widehat{\cal P}_{\nu{\rm I},-{\bf k}},
 \label{eq:Fourier-reality-operators}
\end{equation}
we obtain
\begin{equation}
 \widehat\vartheta_{{\rm I},{\bf k}}^{\dagger}
 =
 \widehat\vartheta_{{\rm I},-{\bf k}}.
 \label{eq:vartheta-Hermitian}
\end{equation}
The inverse Laplacian introduces no additional Hermiticity problem,
because $k^{-2}$ is real and the homogeneous mode $k=0$ is excluded
from the perturbative shift sector.

To specify the ordering of the complete interaction, write the real-space
integrand of Eq.~(\ref{eq:H3eff}) as a sum of individual real monomials
$K_\alpha M_\alpha$, expanding each bracket into its separate terms.
Here $K_\alpha$ includes the sign and all background factors of that
term, while $M_\alpha$ contains its three perturbative factors, including
spatial derivatives where present. Represent $K_\alpha$ by the symmetric
background operator prescribed above and set
\begin{equation}
 \widehat H_{\rm eff}^{(3)}
 :=\frac{N}{2}\sum_\alpha\int d^3x\,
 \left\{\widehat K_\alpha,
 {\cal W}[\widehat M_\alpha]\right\}.
 \label{eq:H3-complete-ordering}
\end{equation}
In this expression ${\cal W}$ permutes the three complete perturbative
factors of each monomial; every shift factor is first defined by
Eq.~(\ref{eq:vartheta-operator-herm}). For example, the term
${\cal B}_{\rm I}\nu{\cal P}_{\nu{\rm I}}^2$ is represented by
$\tfrac12\{\widehat{\cal B}_{\rm I},
{\cal W}[\widehat\nu\widehat{\cal P}_{\nu{\rm I}}
\widehat{\cal P}_{\nu{\rm I}}]\}$.
The outer anticommutator is needed because background coefficients need
not commute with the background operators contained in the perturbative
momentum or shift. With a real classical lapse and a common invariant
operator domain, this prescription gives
\begin{equation}
 {
 \widehat H_{\rm eff}^{(3)\dagger}
 =
 \widehat H_{\rm eff}^{(3)}
 }
 \label{eq:H3-Hermitian}
\end{equation}
formally on the common dense domain of the constituent operators.
Equation~(\ref{eq:H3-Hermitian}) should be understood as a statement of
formal Hermiticity, or symmetry, of the interaction Hamiltonian.  We do
not require here a separate proof of essential self-adjointness, which
would involve a more detailed functional-analytic study and is not
needed for the effective and tree-level calculations considered below.
The effective vertex used here is obtained by replacing the ordered
background factors by their sharply peaked c-number values and neglecting
background moments and correlations. Different orderings can differ by
commutator terms; we do not claim to calculate these quantum corrections
or to prove their smallness for an arbitrary background state.

% We finally consider the auxiliary fiducial cell.  For a noncompact
% spatially flat FLRW geometry one may introduce a cubical coordinate
% cell of volume $V_0$ as an infrared regulator.  In the density
% conventions employed here, the homogeneous quantities
% $a$, $\phi$, $p_\phi$, $\pi_a$, $v=a^3$, $\Omega$, and their effective
% counterparts are local quantities and carry no explicit dependence on
% $V_0$.  In particular, $p_\phi$ and $\pi_a$ denote the homogeneous
% momentum densities used throughout, rather than cell-integrated canonical
% momenta.  The same is true of the local perturbative variables and of the
% Hamiltonian densities.

%\par\endgroup
\subsection{Classical GR and low-curvature limit}
\lb{subsec:GR-limit}

As a further consistency check, we verify explicitly that the cubic
effective Hamiltonian reduces to its classical GR form in the
low-curvature regime.  Define
\begin{equation}
 x\equiv\lambda b,
  \quad |x|\ll1 .
 \label{eq:x-low-curvature}
\end{equation}
Expanding the trigonometric functions gives
\bqn
 \sin^2x
 &=&
 x^2-\frac{x^4}{3}+{\cal O}(x^6),\nb\\
 \sin^2(2x)
 &=&
 4x^2-\frac{16x^4}{3}+{\cal O}(x^6).
 \label{eq:trig-low-curvature}
\eqn
Using Eq.~(\ref{eq:OmegaI}), one then finds
\begin{equation}
 \Omega_{\rm I}^{2}
 =
 v^2b^2
 \left[
 1-
 \left(
 \gamma^2+\frac{4}{3}
 \right)x^2
 +{\cal O}(x^4)
 \right].
 \label{eq:OmegaI-low-curvature}
\end{equation}
On the GR branch the effective Hubble rate follows from the background
Hamiltonian,
\bqn %egin{equation}
 H_{\rm eff}&=&\frac{\sin(2x)}{2\gamma\lambda}
 \left[(1+\gamma^2)\cos(2x)-\gamma^2\right]\nb\\
 &=&\frac{b}{\gamma}\left[1-\left(2\gamma^2+\frac83\right)x^2
 +{\cal O}(x^4)\right].
\eqn %nd{equation}
Thus the canonical variable $\Omega_{\rm can}=vb$ and the effective-rate
variable $\Omega_H=\gamma vH_{\rm eff}$ agree only at leading order:
\[
 \Omega_H=\Omega_{\rm can}\left[1-(2\gamma^2+8/3)x^2+O(x^4)\right].
\]
In the following rate-based comparison, the classical functions
$\mathcal A,\ldots,\mathcal E$, $\mathcal P_\nu$, $F,G$ and $z$ mean their
classical functional forms evaluated at $\Omega_H$. They are comparison
functions, not a redefinition of the canonical variable in Sec.~II.
Expressed in terms of $\Omega_H$, the expansions are
\begin{align}
 \Omega_{\rm I}^{2}
 &=\Omega_H^2\left[1+(3\gamma^2+4)x^2+{\cal O}(x^4)\right],
 \relax\nb\\
 \Lambda_{\rm I}
 &=\Omega_H\left[1+(2\gamma^2+2)x^2+{\cal O}(x^4)\right].
 \label{eq:LambdaI-low-curvature}
\end{align}
It follows that
\bqn %egin{align}
 \frac{1}{\Omega_{\rm I}^{2}}
 &=&\frac{1}{\Omega_H^2}\left[1-(3\gamma^2+4)x^2+{\cal O}(x^4)\right],\nb\\
 \frac{\Lambda_{\rm I}}{\Omega_{\rm I}^{2}}
 &=&\frac{1}{\Omega_H}\left[1-(\gamma^2+2)x^2+{\cal O}(x^4)\right],\nb\\
 \frac{\Lambda_{\rm I}}{\Omega_{\rm I}^{4}}
 &=&\frac{1}{\Omega_H^3}\left[1-(4\gamma^2+6)x^2+{\cal O}(x^4)\right].
 \label{eq:invOmega3-low-curvature}
\eqn %nd{align}
Thus, all three background structures appearing in the effective cubic
Hamiltonian reduce to their classical counterparts:
\begin{equation}
 \frac{1}{\Omega_{\rm I}^{2}}
 \longrightarrow\frac{1}{\Omega_H^2},
  \quad
 \frac{\Lambda_{\rm I}}{\Omega_{\rm I}^{2}}
 \longrightarrow\frac{1}{\Omega_H},
  \quad
 \frac{\Lambda_{\rm I}}{\Omega_{\rm I}^{4}}
 \longrightarrow\frac{1}{\Omega_H^3}.
 \label{eq:inverse-classical-limits}
\end{equation}

For later use, we introduce
\bqn %egin{equation}
 c_1&\equiv&-(\gamma^2+2),
  \quad
 c_2\equiv-(3\gamma^2+4),\nb\\
  \quad
 c_3&\equiv&-(4\gamma^2+6),
 \label{eq:ci-low-curvature}
\eqn %nd{equation}
then the effective cubic coefficients  have the expansions
\begin{align}
 {\cal A}_{\rm I}
 ={}&
 {\cal A}
 +x^2
 \Bigg\{
 c_1
 \left[
 -\frac{3\kappa^2\gamma p_\phi^3}{8v^2\Omega_H}
 +\frac{\kappa\gamma p_\phi}{4\Omega_H}V_{,\phi\phi}
 \right]
 \nonumber\\
 &\hspace{1.20cm}
 +c_3
 \frac{\kappa^3\gamma^3p_\phi^5}
 {16v^2\Omega_H^3}
 \Bigg\}
 +{\cal O}(x^4),
 \relax
 \nb\\
 {\cal B}_{\rm I}
 ={}&
 {\cal B}
 \left[
 1+c_1x^2+{\cal O}(x^4)
 \right],
 \relax
 \nb\\
 {\cal C}_{\rm I}
 ={}&
 {\cal C}
 \left[
 1+c_2x^2+{\cal O}(x^4)
 \right],
 \relax
 \nb\\
 {\cal D}_{\rm I}
 ={}&
 {\cal D}
 \left[
 1+c_1x^2+{\cal O}(x^4)
 \right],
 \relax
 \nb\\
 {\cal E}_{\rm I}
 ={}&
 {\cal E}
 \left[
 1+c_1x^2+{\cal O}(x^4)
 \right].
 \label{eq:EI-low-curvature}
\end{align}
Consequently, we obtain
\bqn %egin{equation}
 && {\cal A}_{\rm I}\rightarrow{\cal A},
 \quad
 {\cal B}_{\rm I}\rightarrow{\cal B},
 \quad
 {\cal C}_{\rm I}\rightarrow{\cal C},\nb\\
 &&
 {\cal D}_{\rm I}\rightarrow{\cal D},
 \quad
 {\cal E}_{\rm I}\rightarrow{\cal E}.
 \label{eq:coeff-classical-limit}
\eqn %nd{equation}

The same check must be made for the composite perturbative variables.
Using Eqs.~(\ref{eq:PnuI-def}) and
(\ref{eq:LambdaI-low-curvature}), we find
\begin{equation}
 {\cal P}_{\nu{\rm I}}
 =
 \pi_\nu
 -
 \frac{\Omega_H}{\gamma v^{2/3}}
 \left(
 1+2(\gamma^2+1)x^2
 +{\cal O}(x^4)
 \right)\nu ,
\end{equation}
and hence
\begin{equation}
 {\cal P}_{\nu{\rm I}}
 =
 {\cal P}_{\nu}
 -
 2(\gamma^2+1)x^2
 \frac{\Omega_H}{\gamma v^{2/3}}\nu
 +{\cal O}(x^4).
 \label{eq:PnuI-low-curvature}
\end{equation}
Therefore, we have
\begin{equation}
 {
 {\cal P}_{\nu{\rm I}}
 \longrightarrow
 {\cal P}_{\nu}.
 }
 \label{eq:Pnu-classical-limit}
\end{equation}

For the shift variable it is useful to write
\begin{equation}
 \vartheta_{{\rm I},{\bf k}}
 =
 \frac{3\kappa}{k^2}
 \left[
 F_{\rm I}\nu_{\bf k}
 +
 G_{\rm I}{\cal P}_{\nu{\rm I},{\bf k}}
 \right],
 \label{eq:vartheta-FG}
\end{equation}
where
\begin{align}
 F_{\rm I}
 ={}&
 \frac{p_\phi}{2v^{2/3}}
 +
 \frac{\gamma v^{4/3}}{6}
 \frac{\Lambda_{\rm I}}{\Omega_{\rm I}^2}V_{,\phi}
 -
 \frac{\kappa\gamma^2p_\phi^3}
 {12v^{2/3}\Omega_{\rm I}^2},
 \nb\\
 G_{\rm I}
 ={}&
 \frac{\gamma p_\phi}{6}
 \frac{\Lambda_{\rm I}}{\Omega_{\rm I}^2}.
\end{align}
Then, Eq.~(\ref{eq:invOmega3-low-curvature})   implies
\begin{align}
 F_{\rm I}
 ={}&
 F
 +x^2
 \left[
 c_1
 \frac{\gamma v^{4/3}}{6\Omega_H}V_{,\phi}
 -
 c_2
 \frac{\kappa\gamma^2p_\phi^3}
 {12v^{2/3}\Omega_H^2}
 \right]
 +{\cal O}(x^4),
 \nb\\
 G_{\rm I}
 ={}&
 G
 \left[
 1+c_1x^2+{\cal O}(x^4)
 \right],
\end{align}
where
\begin{equation}
 F=
 \frac{p_\phi}{2v^{2/3}}
 +
 \frac{\gamma v^{4/3}}{6\Omega_H}V_{,\phi}
 -
 \frac{\kappa\gamma^2p_\phi^3}
 {12v^{2/3}\Omega_H^2},
  \quad
 G=
 \frac{\gamma p_\phi}{6\Omega_H}.
\end{equation}
Together with Eq.~(\ref{eq:PnuI-low-curvature}), this gives
\begin{equation}
 {
 \vartheta_{{\rm I},{\bf k}}
 =
 \vartheta_{\bf k}
 +{\cal O}(x^2),
  \quad
 \vartheta_{{\rm I},{\bf k}}
 \longrightarrow
 \vartheta_{\bf k}.
 }
 \label{eq:vartheta-classical-limit}
\end{equation}

Combining Eqs.~(\ref{eq:coeff-classical-limit}),
(\ref{eq:Pnu-classical-limit}), and
(\ref{eq:vartheta-classical-limit}), every term in
Eq.~(\ref{eq:H3eff}) approaches the corresponding term in the classical cubic
Hamiltonian Eq.~(\ref{eq:H3cl-hybrid}).  Hence
\begin{equation}
 {
 H_{\rm eff}^{(3)}
 =
 H_{\rm cl}^{(3)}
 +{\cal O}\!\left((\lambda b)^2\right),
  \quad
 \lambda b\rightarrow0,
 }
 \label{eq:H3-GR-limit}
\end{equation}
and in particular
\begin{equation}
 {
 H_{\rm eff}^{(3)}
 \longrightarrow
 H_{\rm cl}^{(3)} .
 }
\end{equation}

The curvature conversion has the same limit.  From
Eqs.~(\ref{eq:zI}) and (\ref{eq:CRI}), we find
\begin{equation}
 z_{\rm I}\longrightarrow z,
 \quad
 C_{{\cal R},{\rm I}}\longrightarrow C_{\cal R},
 \quad
 {\cal R}_{\rm eff}\longrightarrow{\cal R}.
 \label{eq:R-eff-GR-limit}
\end{equation}

% This check applies to the GR branch of mLQC-I, for which
% $\lambda b\rightarrow0$.  It should not be confused with the remote
% pre-bounce asymptotic branch, where $b$ approaches a nonzero value and
% the spacetime approaches the characteristic mLQC-I de Sitter regime
% rather than the GR low-curvature branch \cite{Saeed:2024xhk}.

% \begingroup\color{red}
% Sections~\ref{sec:classical} and \ref{sec:effective} therefore provide a
% common canonical foundation for the calculation below: the quadratic and
% cubic Hamiltonians are expressed in the same hybrid variables, the mLQC-I
% regularization is fixed before the effective background constraint is used,
% and both the interaction Hamiltonian and the late-time curvature conversion
% recover their classical GR forms on the expanding low-curvature branch.

% For a comparison at fixed canonical momentum, the corresponding inverse structures are
% \[
%  \frac{\Lambda_{\rm I}}{\Omega_{\rm I}^2}
%  =\frac{1+(\gamma^2+2/3)x^2+O(x^4)}{\Omega_{\rm can}},
% \]
% \[
%  \frac1{\Omega_{\rm I}^2}
%  =\frac{1+(\gamma^2+4/3)x^2+O(x^4)}{\Omega_{\rm can}^2},
% \]
% \[
%  \frac{\Lambda_{\rm I}}{\Omega_{\rm I}^4}
%  =\frac{1+(2\gamma^2+2)x^2+O(x^4)}{\Omega_{\rm can}^3}.
% \]
% These positive subleading coefficients and the rate-based coefficients above
% are consistent after the change of comparison variable. 

% \par\endgroup
\section{Correlation functions and perturbativity diagnostic}
\renewcommand{\theequation}{4.\arabic{equation}}
\setcounter{equation}{0}
\label{sec:correlation-functions}

\subsection{Exact finite-momentum cubic vertex}
For a momentum triangle
\begin{equation}
 \mathbf{k}_1+\mathbf{k}_2+\mathbf{k}_3=0,
\end{equation}
we define
\begin{equation}
 d_{ij}\equiv\mathbf{k}_i\cdot\mathbf{k}_j,
 \quad
 Q_{ij}\equiv d_{ij}^2-k_i^2k_j^2.
\end{equation}
For each leg we introduce the linear interaction-picture quantities
\begin{equation}
 u_i\equiv u_{k_i},
 \quad
 p_i\equiv u_i'-\Gamma_{\rm I}u_i,
 \quad
 y_i\equiv F_{\rm I}u_i+G_{\rm I}p_i.
 \label{eq:upy}
\end{equation}
Then, we have
\begin{equation}
 \vartheta_i=\frac{3\kappa}{k_i^2}y_i.
\end{equation}

After complete symmetrization of the three legs, it is numerically convenient
to use the vertex with no divisions by the mode functions:
\begin{align}
 {\mathscr V}_{123}
 ={}&
 {\cal A}_{\rm I}u_1u_2u_3
 \nonumber\\
 &+
 \frac{{\cal B}_{\rm I}}{3}
 \left(
 u_1p_2p_3+u_2p_3p_1+u_3p_1p_2
 \right)
 \nonumber\\
 &-
 \frac{{\cal C}_{\rm I}}{3}
 \left(
 u_1u_2p_3+u_2u_3p_1+u_3u_1p_2
 \right)
 \nonumber\\
 &+
 \frac{{\cal B}_{\rm I}}{6}
 (k_1^2+k_2^2+k_3^2)u_1u_2u_3
 \nonumber\\
 &+
 \frac{\kappa{\cal D}_{\rm I}}{2}
 \left(
 y_1u_2u_3+y_2u_3u_1+y_3u_1u_2
 \right)
 \nonumber\\
 &-
 \frac{\kappa}{2a^2}
 \sum_{\rm cyc}
 \frac{y_i}{k_i^2}
 \left(
 d_{ki}p_j u_k+d_{ji}p_k u_j
 \right)
 \nonumber\\
 &+
 3\kappa^2{\cal E}_{\rm I}
 \sum_{\rm cyc}
 \frac{Q_{jk}}{k_j^2k_k^2}
 u_i y_jy_k,
 \label{eq:Vexact}
\end{align}
where $(i,j,k)$ runs over the cyclic permutations of $(1,2,3)$.
Equation~(\ref{eq:Vexact}) is exact at finite momentum; no ultraviolet
approximation has been made.
Under a constant rephasing $u_i\mapsto e^{i\theta_i}u_i$, both $p_i$
and $y_i$ acquire the same phase as $u_i$. Every term in
${\mathscr V}_{123}$ therefore acquires the common factor
$e^{i(\theta_1+\theta_2+\theta_3)}$. Its conjugate in
Eq.~(\ref{eq:Bnu}) cancels the phase of the three external modes, so
the analytic bispectrum is invariant. 
%This identity supplies a numericalimplementation test. 
% Representative numerical checks are reported in
% Appendix~\ref{app:validation-bounce}; their residuals apply to the tested
% cases rather than supplying a uniform tolerance for all triangles.

\subsection{In-in bispectrum of the hybrid MS variable}
We define
\begin{equation}
 \left\langle
 \hat\nu_{\mathbf{k}_1}
 \hat\nu_{\mathbf{k}_2}
 \hat\nu_{\mathbf{k}_3}
 \right\rangle
 =
 (2\pi)^3\delta^{(3)}
 (\mathbf{k}_1+\mathbf{k}_2+\mathbf{k}_3)
 B_\nu(k_1,k_2,k_3).
 \label{eq:Bnu-def}
\end{equation}
For a Gaussian initial state the free three-point function vanishes.  To
first order in the cubic interaction, we have
\begin{align}
 \left\langle\nu^3(\eta)\right\rangle
 =
 -i\int_{\eta_0}^{\eta}\dd\eta'\,
 \left\langle
 \left[
 \nu^3(\eta),H_I^{(3)}(\eta')
 \right]
 \right\rangle.
 \label{eq:inin}
\end{align}
Because the completely symmetrized Hamiltonian in Eq.~(\ref{eq:H3eff}) carries
no explicit factor $1/3!$, the three external fields can be contracted with
the three fields at the vertex in $3!=6$ equivalent ways.  Consequently, we find
\begin{align}
 B_\nu(k_1,k_2,k_3;\eta)
 ={}&
 12\,{\rm Im}
 \Bigg[
 u_1(\eta)u_2(\eta)u_3(\eta)
 \nonumber\\
 &\times
 \int_{\eta_0}^{\eta}\dd\eta'\,
 a(\eta'){\mathscr V}_{123}^{*}(\eta')
 \Bigg].
 \label{eq:Bnu}
\end{align}
It is useful to define
\begin{equation}
 I_{123}'(\eta)
 =
 6a(\eta){\mathscr V}_{123}^{*}(\eta),
 \quad
 I_{123}(\eta_0)=0,
 \label{eq:Iode}
\end{equation}
so that
\begin{equation}
 B_\nu(k_1,k_2,k_3;\eta)
 =
 2\,{\rm Im}
 \left[
 u_1(\eta)u_2(\eta)u_3(\eta)I_{123}(\eta)
 \right].
 \label{eq:Bnu-I}
\end{equation}
Equation~(\ref{eq:Iode}) allows the in-in integral to be evolved
simultaneously with the three linear mode equations, which is particularly
convenient for the numerical calculation.

\subsection{Late-time conversion to the curvature perturbation}

The classical nonlinear relation was given in Sec.~\ref{subsec:classical-R},
and its minimal mLQC-I continuation in Sec.~\ref{subsec:R-eff}.  We evaluate
the correlation functions only on a late super-Hubble surface in the
expanding classical regime, where Eq.~(\ref{eq:R-eff-GR-limit}) applies.
Accordingly, with
$z_s=a\dot\phi/H=z$ on that surface, the local relation becomes
\begin{equation}
 \calR
 =
 q+C_{\calR}q^2+\cdots,
 \quad
 q\equiv-\frac{\nu}{z_s},
 \label{eq:R-nonlinear}
\end{equation}
where
\begin{equation}
 C_{\calR}
 =
 -\frac32
 -\frac{V_{,\phi}}{2H\dot\phi}
 +\frac{\kappa\dot\phi^2}{4H^2}.
 \label{eq:CR}
\end{equation}
%\begingroup\color{blue}
Spatial-gradient and time-derivative quadratic terms are omitted in
Eq.~(\ref{eq:R-nonlinear}) as a local late-time approximation
\cite{Agullo:2017eyh}. 
%\par\endgroup
  On the late classical branch, the background
Klein--Gordon equation gives
$V_{,\phi}=-\ddot\phi-3H\dot\phi$.  Hence
\begin{equation}
 C_{\calR}
 =
 \frac{1}{2}\frac{\ddot\phi}{H\dot\phi}
 +\frac{\epsilon_H}{2},
 \quad
 \epsilon_H\equiv-\frac{\dot H}{H^2}
 =\frac{\kappa\dot\phi^2}{2H^2},
 \label{eq:CR-sr}
\end{equation}
where the final equality for $\epsilon_H$ applies in the classical GR
regime.
% \begingroup\color{blue}\label{rv18}
%  With $\delta_\phi\equiv\ddot\phi/(H\dot\phi)$, the exact late-GR
% identity is $C_{\cal R}=(\epsilon_H+\delta_\phi)/2$. Approximating it by
% $\epsilon_H/2$ requires $|\delta_\phi|\ll|\epsilon_H|$, not merely
% $|\delta_\phi|\ll1$. A relative tolerance $\tau$ with respect to
% $\epsilon_H/2$ requires $|\delta_\phi|\leq\tau|\epsilon_H|$; with respect
% to the exact coefficient it requires
% $|\delta_\phi|\leq\tau|\epsilon_H+\delta_\phi|$.
%
% \par\endgroup
In Fourier space,
\begin{equation}
 \calR_{\mathbf{k}}
 =
 -\frac{\nu_{\mathbf{k}}}{z_s}
 +
 \frac{C_{\calR}}{z_s^2}
 \int_{\mathbf{p}}
 \nu_{\mathbf{p}}\nu_{\mathbf{k}-\mathbf{p}}
 +\cdots .
 \label{eq:R-Fourier}
\end{equation}
Because the linear term in Eq.~(\ref{eq:R-Fourier}) carries a minus sign,
the cross term between the linear and quadratic pieces of $\calR$ carries a
minus sign in the retained two-point correction.  The Wick term is
quadratic in $C_{\calR}$ and is unaffected by this sign.

\subsection{Retained nonlinear-conversion correction to the curvature power spectrum}
We define
\bqn %egin{equation}
 \left\langle
 \hat\calR_{\mathbf{k}}
 \hat\calR_{\mathbf{k}'}
 \right\rangle
 &=&
 (2\pi)^3\delta^{(3)}(\mathbf{k}+\mathbf{k}')
 \frac{2\pi^2}{k^3}\nb\\
 && \times
 \left[
 \calP_{\calR}^{(0)}(k)+\Delta\calP_{\calR}(k)
 \right].
 \label{eq:PR-def}
\eqn %nd{equation}
At leading order, we have
\begin{equation}
 \calP_{\calR}^{(0)}(k)
 =
 \frac{k^3}{2\pi^2z_s^2}|u_k|^2.
 \label{eq:PR0}
\end{equation}

The correction linear in $H_I^{(3)}$ to the ordinary two-point function
$\langle\nu\nu\rangle$ vanishes for a Gaussian interaction-picture state,
because it contains an odd number of free fields.  At the order retained
here, the nonlinear conversion in Eq.~(\ref{eq:R-Fourier}) therefore gives
two contributions: a term involving the tree-level bispectrum and a Gaussian
four-point (Wick) term.  One obtains
\begin{align}
 \Delta\calP_{\calR}(k)
 ={}&
 -\frac{k^3}{\pi^2}
 \frac{C_{\calR}}{z_s^3}
 \int\frac{\dd^3p}{(2\pi)^3}
 B_\nu
 \left(
 \mathbf{k},\mathbf{p},-\mathbf{k}-\mathbf{p}
 \right)
 \nonumber\\
 &+
 \frac{k^3}{\pi^2}
 \frac{C_{\calR}^2}{z_s^4}
 \int\frac{\dd^3p}{(2\pi)^3}
 |u_p|^2|u_{|\mathbf{k}-\mathbf{p}|}|^2 .
 \label{eq:DeltaPR}
\end{align}
All conversion factors in Eq.~(\ref{eq:DeltaPR}) are evaluated at the
late-time super-Hubble evaluation surface.

For numerical purposes, we define
\begin{align}
 I_B(k)
 &\equiv
 \int\frac{\dd^3p}{(2\pi)^3}
 B_\nu(\mathbf{k},\mathbf{p},-\mathbf{k}-\mathbf{p}),
 \nb\\
 I_W(k)
 &\equiv
 \int\frac{\dd^3p}{(2\pi)^3}
 |u_p|^2|u_{|\mathbf{k}-\mathbf{p}|}|^2 .
\end{align}

With these definitions, we find
\begin{equation}
 \frac{\Delta\calP_{\calR}}
      {\calP_{\calR}^{(0)}}
 =
 -\frac{2C_{\calR}}{z_s|u_k|^2}I_B(k)
 +
 \frac{2C_{\calR}^2}{z_s^2|u_k|^2}I_W(k).
 \label{eq:ratio}
\end{equation}

Equation~(\ref{eq:ratio}) directly measures the nonlinear correction to
the curvature two-point function retained in the present calculation.  A
closely related correlation-function criterion was used in standard LQC in
Ref.~\cite{Agullo:2017eyh}.  Genuine $O([H_I^{(3)}]^2)$ corrections to
$\langle\nu\nu\rangle$ are not included here and are of the same perturbative order as the retained conversion terms and
require a separate calculation. Quartic interactions and the cubic curvature
conversion may also contribute at that order \cite{Saeed:2026xxx}.

More explicitly, write the curvature conversion as
${\cal R}=q+{\cal R}_2+{\cal R}_3+\cdots$, where
${\cal R}_n$ has degree $n$ in the linear perturbations. At leading
Gaussian order the power spectrum is proportional to one power of the
linear covariance. The next connected two-point order, proportional to
two powers of that covariance, generally includes
\begin{equation}
 \begin{split}
 &\langle{\cal R}_2{\cal R}_2\rangle_0,
 \quad 2\langle q{\cal R}_2\rangle_{H_3},
 \quad 2\langle q{\cal R}_3\rangle_0,\\
 &\langle qq\rangle_{H_3^2},
 \quad \langle qq\rangle_{H_4}.
 \end{split}
 \label{eq:two-point-power-counting}
\end{equation}
The subscripts specify the interaction insertions, with $0$ denoting
free Gaussian contractions. Equation~(\ref{eq:DeltaPR}) retains only
the first two structures, using the local late-time approximation
${\cal R}_2=C_{\cal R}q^2$. Interaction-insertion order is therefore
distinct from the final order in the covariance expansion. No hierarchy
suppressing the other structures in Eq.~(\ref{eq:two-point-power-counting})
is established by the retained diagnostic alone.

\subsection{Correlation-function perturbativity criterion}

It is useful to separate the two contributions in
Eq.~(\ref{eq:ratio}) by defining
\begin{align}
 \delta_B(k)
 &\equiv
 -\frac{2C_{\calR}}{z_s|u_k|^2}I_B(k),
 \label{eq:deltaB}
 \\
 \delta_W(k)
 &\equiv
 \frac{2C_{\calR}^2}{z_s^2|u_k|^2}I_W(k).
 \label{eq:deltaW}
\end{align}
The perturbativity diagnostic is then
\begin{equation}
 \Xi(k)
 =
 \left|\delta_B(k)+\delta_W(k)\right|.
 \label{eq:Xi-components}
\end{equation}
In the numerical analysis we shall monitor not only $\Xi$, but also
$|\delta_B|$ and $|\delta_W|$ separately.  This prevents an accidental
cancellation between the bispectrum and Wick terms from being mistaken for
uniform perturbative suppression.

For the numerical calculation it is also useful to introduce a conservative
quantity that does not rely on cancellations within the oscillatory
bispectrum convolution.  From the triangle inequality we define
\begin{align}
 |\delta_B(k)|_{\rm abs}
 \equiv{}&
 \frac{|C_{\calR}|}
 {2\pi^2 k\,|z_s|\,|u_k|^2}
 \int_{k_{\rm IR}}^{k_{\rm UV}}\dd p\,p
 \nonumber\\
 &\times
 \int_{q_-}^{q_+}\dd q\,q\,
 |B_\nu(k,p,q)|,
 \label{eq:deltaB-abs}
\end{align}
where the finite triangle domain is specified below.  The associated
conservative perturbativity diagnostic is
\begin{equation}
 \Xi_{\rm abs}(k)
 \equiv
 |\delta_B(k)|_{\rm abs}+|\delta_W(k)|.
 \label{eq:Xi-abs}
\end{equation}
By construction, we have
\begin{equation}
 \Xi(k)\leq \Xi_{\rm abs}(k).
 \label{eq:Xi-bound}
\end{equation}
Thus, $\Xi_{\rm abs}\ll1$ is a sufficient condition for perturbative
smallness of the correction retained in the present correlation-function test, even when
the signed bispectrum integral is difficult to resolve because of rapid
momentum-space oscillations.

\subsection{Reduction of the momentum convolutions and domain of validity}
\label{subsec:convolution-domain}

For an isotropic background the three-dimensional momentum convolutions can
be reduced to two-dimensional integrals over momentum triangles.  Choose the
external momentum $\mathbf{k}$ along the $z$ axis and write
\begin{align}
 p&\equiv|\mathbf{p}|,
 \quad
 \mu\equiv\hat{\mathbf{k}}\cdot\hat{\mathbf{p}},
 \nonumber\\
 q&\equiv|\mathbf{k}+\mathbf{p}|
 =\sqrt{k^2+p^2+2kp\mu}.
 \label{eq:p-mu-q}
\end{align}
Rotational invariance gives
\begin{equation}
 \frac{\dd^3p}{(2\pi)^3}
 =
 \frac{p^2\,\dd p\,\dd\mu}{4\pi^2},
 \label{eq:measure-pmu}
\end{equation}
and, since $q\,\dd q=kp\,\dd\mu$, the angular range
$-1\leq\mu\leq1$ is equivalent to the triangle condition
\begin{equation}
 |k-p|\leq q\leq k+p.
 \label{eq:triangle-domain}
\end{equation}
The bispectrum convolution can therefore be written as
\begin{equation}
 I_B(k)
 =
 \frac{1}{4\pi^2 k}
 \int_0^\infty \dd p\,p
 \int_{|k-p|}^{k+p}\dd q\,q\,
 B_\nu(k,p,q).
 \label{eq:IB-2d-formal}
\end{equation}
For the Wick contribution one may make the change of variables
$\mathbf{p}\rightarrow-\mathbf{p}$, after which the same triangle variables
can be used.  Hence
\begin{equation}
 I_W(k)
 =
 \frac{1}{4\pi^2 k}
 \int_0^\infty \dd p\,p
 \int_{|k-p|}^{k+p}\dd q\,q\,
 |u_p|^2|u_q|^2 .
 \label{eq:IW-2d-formal}
\end{equation}
Equations~(\ref{eq:IB-2d-formal}) and
(\ref{eq:IW-2d-formal}) are the formal triangle representations of the
convolutions entering the present perturbativity test.

There is an important restriction on how these expressions are used in the
present calculation.  The nonlinear conversion
Eq.~(\ref{eq:R-nonlinear}) was simplified by neglecting spatial-gradient and
time-derivative terms on a late-time hypersurface for which the relevant
modes are well outside the Hubble radius.  It would therefore be inconsistent
to combine this super-Hubble approximation with an unrestricted integration
over arbitrarily large internal momenta, for which the omitted derivative
terms need not be negligible.  In the numerical implementation we instead
restrict all three legs of the momentum triangle to
\begin{equation}
 k_{\rm IR}\leq k,p,q\leq k_{\rm UV},
 \quad
 k_{\rm UV}\ll a_fH_f,
 \label{eq:convolution-window}
\end{equation}
where $\eta_f$ is the late-time evaluation surface.  The internal integration
limits are then
\begin{align}
 q_-(k,p)
 &\equiv
 \max\left\{|k-p|,k_{\rm IR}\right\},
 \nb\\
 q_+(k,p)
 &\equiv
 \min\left\{k+p,k_{\rm UV}\right\},
 \label{eq:qpm}
\end{align}
with no contribution whenever $q_+\leq q_-$.  The numerical convolutions are
therefore
\begin{align}
 I_B(k)
 ={}&
 \frac{1}{4\pi^2 k}
 \int_{k_{\rm IR}}^{k_{\rm UV}}\dd p\,p
 \int_{q_-}^{q_+}\dd q\,q\,
 B_\nu(k,p,q),
 \relax
 \nb\\
 I_W(k)
 ={}&
 \frac{1}{4\pi^2 k}
 \int_{k_{\rm IR}}^{k_{\rm UV}}\dd p\,p
 \int_{q_-}^{q_+}\dd q\,q\,
 |u_p|^2|u_q|^2 .
 \label{eq:IW-2d-window}
\end{align}
Substitution into Eqs.~(\ref{eq:deltaB}) and (\ref{eq:deltaW}) gives the
forms used directly in the numerical calculation,
\begin{align}
 \delta_B(k)
 ={}&
 -\frac{C_{\calR}}
 {2\pi^2 k\,z_s|u_k|^2}
 \int_{k_{\rm IR}}^{k_{\rm UV}}\dd p\,p
 \int_{q_-}^{q_+}\dd q\,q\,
 B_\nu(k,p,q),
 \label{eq:deltaB-2d}
 \\
 \delta_W(k)
 ={}&
 \frac{C_{\calR}^2}
 {2\pi^2 k\,z_s^2|u_k|^2}
 \int_{k_{\rm IR}}^{k_{\rm UV}}\dd p\,p
 \int_{q_-}^{q_+}\dd q\,q\,
 |u_p|^2|u_q|^2 .
 \label{eq:deltaW-2d}
\end{align}

% \begingroup\color{blue}
% The boundaries in Eq.~(\ref{eq:convolution-window}) define the finite-domain
% observable estimated here. Varying them tests domain sensitivity.
% \par\endgroup
  Extending the convolution to
arbitrarily large internal momentum would require retaining the complete
second-order field redefinition rather than its super-Hubble reduction and,
in general, treating the ultraviolet definition of the resulting composite
operators consistently.  We therefore do not interpret dependence on an
uncontrolled large-$p$ extension of Eqs.~(\ref{eq:IB-2d-formal}) and
(\ref{eq:IW-2d-formal}) as part of the correlation-function test performed
here.

For orientation, the size of the two terms can be estimated on the
late-time slow-roll surface.  There,
$C_{\calR}=O(\epsilon_H,\ddot\phi/(H\dot\phi))$.  For comparable momenta
$k_1\sim k_2\sim k_3\sim k$, after factoring out the standard momentum
dependence of the bispectrum, its dimensionless amplitude scales schematically
as
\begin{equation}
 k^6 B_{\calR}
 \sim
 f_{\rm NL}
 \left(\calP_{\calR}^{(0)}\right)^2 .
\end{equation}
Consequently, up to momentum-shape factors and the convolution measure, the
two contributions scale parametrically as
\begin{align}
 \delta_B
 &\sim
 O\!\left(
 C_{\cal R} f_{\rm NL}\calP_{\calR}^{(0)}
 \right),
 \label{eq:scaling-B}
 \\
 \delta_W
 &\sim
 O\!\left(
C_{\cal R}^2\calP_{\calR}^{(0)}
 \right).
 \label{eq:scaling-W}
\end{align}
Thus, even a comparatively large bispectrum need not invalidate the
correlation-function expansion, because the correction to the two-point
function contains an additional power of the small scalar power spectrum.
The same suppression mechanism was emphasized for standard LQC in
Ref.~\cite{Agullo:2017eyh}.  Equations~(\ref{eq:scaling-B}) and
(\ref{eq:scaling-W}) are only order-of-magnitude estimates; all numerical results below are reported from the full finite-momentum expression (\ref{eq:ratio}), without replacing the bispectrum by an $f_{\rm NL}$
ansatz.

\section{Initial states and correlation-function perturbativity}
\renewcommand{\theequation}{5.\arabic{equation}}
\setcounter{equation}{0}
\label{sec:states-results}

We now evaluate the correlation-function diagnostic for two physically
distinct initial-state prescriptions while keeping the background dynamics,
cubic Hamiltonian, late-time conversion, and momentum-domain prescription
fixed.

\subsection{Common numerical strategy}
For either prescription, the linear mode equation is solved once on the
required momentum grid and the histories $u_k(\eta)$ and $u_k'(\eta)$ are
stored.  For each momentum triangle we then construct the exact vertex
$\mathscr V_{123}$ from this mode bank and integrate Eq.~(\ref{eq:Iode}) to
the late super-Hubble evaluation surface.  The bispectrum follows from
Eq.~(\ref{eq:Bnu-I}), after which the momentum convolutions in
Eq.~(\ref{eq:IW-2d-window}) are performed.

%Three numerical checks are essential: convergence with respect to the time
%discretization of the in-in integral, convergence of the momentum
%convolution, and independence of the late evaluation time after the relevant
%modes have frozen out.  
The bispectrum convolution is highly oscillatory in
the internal momenta.  %\begingroup\color{blue}
The reported implementation uses a triangle-aligned composite grid and
compares two spacings. 
%\par\endgroup
  In parallel we evaluate
$\Xi_{\rm abs}$, which is insensitive to cancellations between neighboring
oscillations and therefore provides a conservative stability diagnostic.

The control parameters $k_{\rm IR}$ and $k_{\rm UV}$ in
Eq.~(\ref{eq:convolution-window}) are varied explicitly while maintaining the
super-Hubble condition $k_{\rm UV}\ll a_fH_f$ required by the late-time field
redefinition.  Strict cutoff independence is not assumed; when a tested
window has not reached a plateau we quote the result as a finite-window estimate.

An independently implemented alternative expands the exact vertex into a
finite sum of separable radial contractions and uses locally validated
oscillatory time quadrature; see Appendix~\ref{app:independent-validation}.
Its explicit candidate state and fixed control window distinguish these
checks from a reproduction of every numerical table below. Spatial,
temporal, interpolation, and initial-state tests are kept separate.

\subsection{Bounce-defined initial state}
\label{subsec:bounce-results}

\subsubsection{Initial state and numerical setup}
The first initial-state prescription is specified directly at the bounce,
\begin{equation}
 \eta_0=\eta_B.
\end{equation}
For the finite momentum windows considered below, the linear mode functions
are initialized in the instantaneous quadratic ground state at the bounce.
A smooth fourth-order adiabatic completion would be required if this
prescription were extended arbitrarily far into the ultraviolet, but it is
not needed for the finite-window results reported here.
Explicit instantaneous data are
$u_{kB}=(2\omega_{kB})^{-1/2}$ and $u'_{kB}=-i\omega_{kB}u_{kB}$, with
$\omega_{kB}^2=k^2+m_{\rm eff}^2(\eta_B)$.

For definiteness we use the Starobinsky potential
\begin{equation}
 V(\phi)
 =
 \frac{3m_\star^2}{32\pi G}
 \left(1-e^{-\sqrt{16\pi G/3}\,\phi}\right)^2,
 \label{eq:V-Starobinsky-num}
\end{equation}
with
\begin{equation}
 m_\star=3.08\times10^{-6}m_{\rm P},\quad
 \gamma=0.2375,
\end{equation}
and bounce data
\begin{equation}
 a_B=1,\quad
 \phi_B=-1.09\,m_{\rm P},\quad
 \dot\phi_B>0.
\end{equation}
For the enlarged-window scan below we evaluate the curvature perturbation at
$N\equiv\ln a=18$ and integrate over
\begin{equation}
 0.01\leq p/m_{\rm P},q/m_{\rm P}\leq2.05.
 \label{eq:num-window-main}
\end{equation}
At this time $aH\simeq1.01\times10^2m_{\rm P}$, so that even the largest
internal momentum satisfies $k_{\rm UV}/(aH)\simeq2.0\times10^{-2}$.
All internal momenta in this window are therefore super-Hubble. 
% This
% supports a gradient expansion, but does not by itself bound the omitted
% derivative and shift contributions relative to the small retained local
% correction. Such a relative error requires an explicit estimate from the
% full conversion or additional late-time tests.

% The mode equation preserves a canonical Wronskian, and the finite-momentum
% cubic vertex is symmetric under permutations of its three legs. Quantitative checks in Appendix~\ref{app:validation-bounce} report approximately
% $4\times10^{-15}$ relative vertex-permutation agreement and a fractional
% power change below $1.1\times10^{-7}$ in the stated solver comparison.
% }

\begin{figure}[!tbp]

 \centering
 \includegraphics[width=\columnwidth]{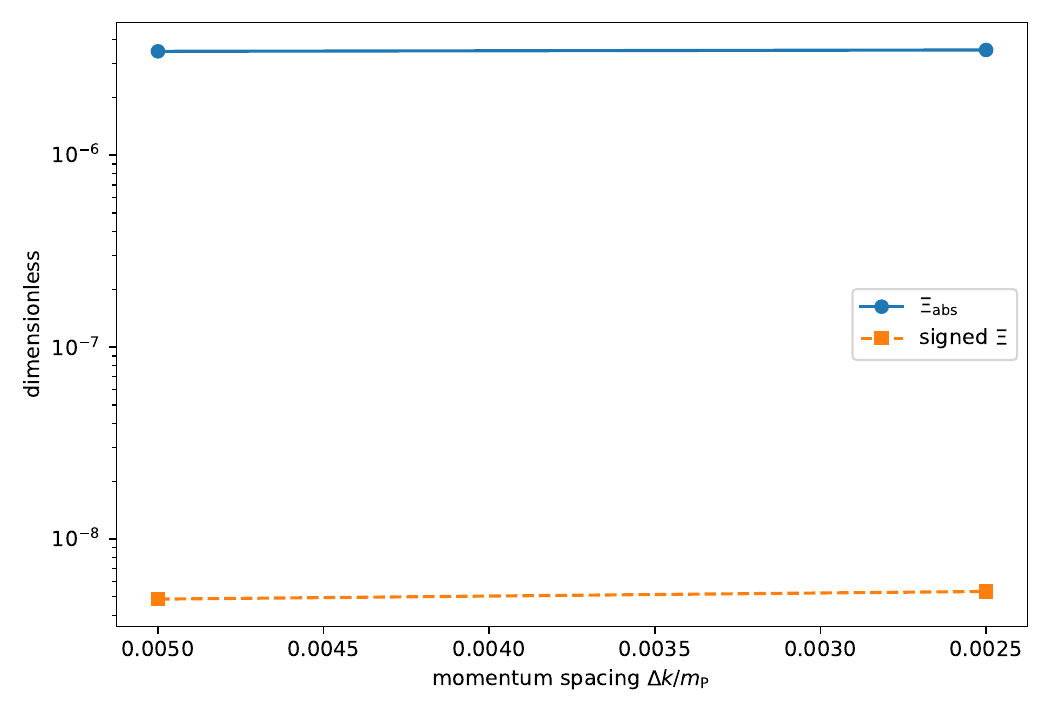}
 \caption{Momentum-resolution test at $k=0.3m_{\rm P}$ for the
 bounce-defined state on the control window specified in the text.
The two quoted estimates of $\Xi_{\rm abs}$ differ by about $1.44\%$.
 }
 \label{fig:resolution-convergence}
\end{figure}

  The momentum
integral is evaluated on a triangle-aligned composite grid.  A dedicated
resolution test at $k=0.3m_{\rm P}$ gives
\begin{align}
 \Xi_{\rm abs}
 &=3.48\times10^{-6},
 &&\Delta k=0.005m_{\rm P},
 \nonumber\\
 &=3.53\times10^{-6},
 &&\Delta k=0.0025m_{\rm P},
 \label{eq:resolution-values}
\end{align}
for the smaller control window
$0.05\leq p/m_{\rm P},q/m_{\rm P}\leq1.05$.  The change based on the rounded values is about $1.44\%$.
Figure~\ref{fig:resolution-convergence} displays this refinement test.
%\begingroup\color{blue}
The quoted $\Delta k_t$ was measured at $k_t=3,8,16,24m_{\rm P}$,
whereas this control window has $k_t\leq2.4m_{\rm P}$.

%\par\endgroup

\subsubsection{Finite-window absolute-integrand diagnostic}
\begin{figure}[!tbp]

 \centering
 \includegraphics[width=\columnwidth]{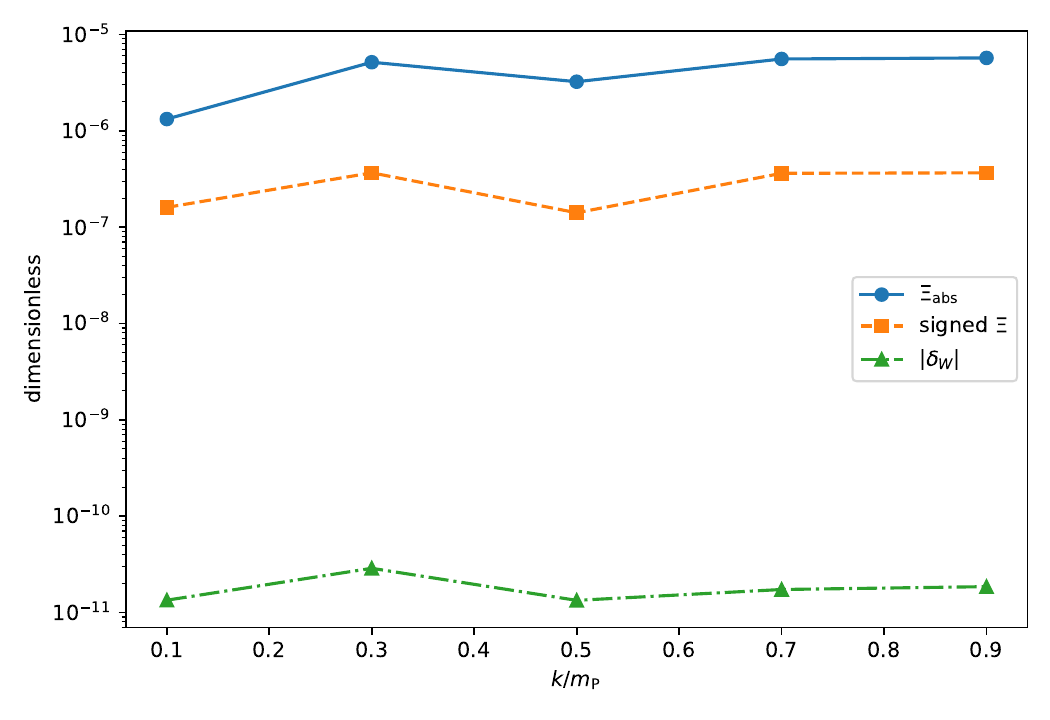}
 \caption{Enlarged-window scan of the perturbativity diagnostics for
 the bounce-defined state at $N=18$.  The conservative
 $\Xi_{\rm abs}$ does not use cancellations in the oscillatory bispectrum
 convolution and is estimated below $6\times10^{-6}$ at the five sampled
 momenta.  The magnitude $\Xi$ of the signed correction is smaller but is more sensitive to momentum resolution. The legend\textquotesingle s ``signed $\Xi$\textquotesingle\textquotesingle\ denotes this magnitude.}
 \label{fig:Xi-scan}
\end{figure}

The enlarged-window scan is summarized in Table~\ref{tab:Xi-scan}.  Over the
five sampled external momenta in $0.1\leq k/m_{\rm P}\leq0.9$, the
absolute-integrand diagnostic is estimated at the few-$10^{-6}$ level.  The Wick term is much smaller, of order
$10^{-11}$, so the diagnostic is dominated by the integral of the absolute value of the
bispectrum.  The same enlarged-window scan is shown graphically
in Fig.~\ref{fig:Xi-scan}, which shows estimates of $\Xi_{\rm abs}$ below
$6\times10^{-6}$ at the five sampled momenta while the Wick
contribution is negligible on this scale.

\begin{table}[!tbp]
 \caption{Enlarged-window perturbativity scan at $N=18$ for the
 bounce-defined state.  The internal range is
 $0.01\leq p/m_{\rm P},q/m_{\rm P}\leq2.05$.}
 \label{tab:Xi-scan}
 \begin{ruledtabular}
 \begin{tabular}{cccc}
 $k/m_{\rm P}$ & $\calP_{\calR}^{(0)}$ & $\Xi_{\rm abs}$ & $\Xi$ \\
 \hline
 0.10 & $4.40\times10^{-8}$ & $1.32\times10^{-6}$ & $1.60\times10^{-7}$ \\
 0.30 & $1.21\times10^{-8}$ & $5.15\times10^{-6}$ & $3.67\times10^{-7}$ \\
 0.50 & $3.25\times10^{-8}$ & $3.23\times10^{-6}$ & $1.41\times10^{-7}$ \\
 0.70 & $2.03\times10^{-8}$ & $5.57\times10^{-6}$ & $3.62\times10^{-7}$ \\
 0.90 & $1.83\times10^{-8}$ & $5.71\times10^{-6}$ & $3.66\times10^{-7}$
 \end{tabular}
 \end{ruledtabular}
\end{table}

The rapid sign oscillations of $B_{\nu}(k,p,q)$ make the signed bispectrum
integral more demanding numerically than the absolute-integrand diagnostic.  Its
precise residual value therefore should not yet be interpreted as a
high-accuracy prediction.  For the exact integrals on the same finite domain, Eq.~(\ref{eq:Xi-bound}) gives
\begin{equation}
 \Xi(k)\leq\Xi_{\rm abs}(k).
 \label{eq:first-main-bound}
\end{equation}
The numerical estimates of $\Xi_{\rm abs}$ at the five sampled momenta are below $6\times10^{-6}$.

\subsubsection{Late-time and momentum-window checks}
At fixed $k=0.3m_{\rm P}$ and for the control window
$0.05\leq p/m_{\rm P},q/m_{\rm P}\leq1.05$, the conservative result changes
only mildly between $N=16$ and $N=18$,
\begin{align}
 \Xi_{\rm abs}(N=16)&=3.55\times10^{-6},\nonumber\\
 \Xi_{\rm abs}(N=17)&=3.57\times10^{-6},\nonumber\\
 \Xi_{\rm abs}(N=18)&=3.60\times10^{-6}.
 \label{eq:late-time-check}
\end{align}
% The value $3.60\times10^{-6}$ in this late-time test differs from the
% $3.53\times10^{-6}$ quoted in the resolution test on the same stated
% window, a $1.983\%$ difference relative to $3.53\times10^{-6}$.

These values indicate mild variation in
this integrated estimate on the
tested surfaces. However, they do not by themselves establish pointwise bispectrum
stability or bound the omitted derivative terms.

The dependence on the finite momentum window is more important.  With
$k_{\rm IR}=0.05m_{\rm P}$, increasing $k_{\rm UV}$ from
$0.65m_{\rm P}$ to $2.05m_{\rm P}$ changes the estimated diagnostic at
$k=0.3m_{\rm P}$ from approximately $2.94\times10^{-6}$ to
$3.93\times10^{-6}$.  At fixed $k_{\rm UV}=1.05m_{\rm P}$, lowering
$k_{\rm IR}$ from $0.10m_{\rm P}$ to $0.005m_{\rm P}$ raises it from
approximately $2.74\times10^{-6}$ to $5.36\times10^{-6}$.  The ultraviolet
accumulation is flattening over the tested range, whereas the infrared side
has not yet reached a clear cutoff-independent plateau.  We therefore regard
the tabulated absolute-integrand values only as finite-window estimates.
% They do not establish a regulator-independent infrared limit or a numerical
% error enclosure.

\subsection{Remote-contracting de Sitter initial state}
\label{subsec:remote-results}

\subsubsection{State selection and contracting-branch adiabaticity}
\label{subsec:contracting-adiabaticity}

The second prescription is the de Sitter vacuum in the remote contracting
branch.  In the asymptotic pre-de Sitter regime the positive-frequency mode
can be written as
\begin{equation}
 u_k^{\rm dS}(\eta)
 =
 \sqrt{\frac{\pi}{4k}}\,\sqrt{k\eta}\,
 H_{\mu}^{(2)}(k\eta),
 \quad
 \mu=\sqrt{\frac14-\epsilon_\Lambda},
 \label{eq:remote-dS-state}
\end{equation}
with the coefficient of $H_\mu^{(1)}$ set to zero.  For the representative
potential and background initial data used below,
$\epsilon_\Lambda=O(10^{-5})$. 

We impose these data at a finite time sufficiently far in the contracting phase
and assess the sensitivity to moving that surface farther into the past.
Appendix~\ref{app:validation-remote-result} supplies a finite-start
signed-convolution comparison for an explicitly specified independent
candidate. 
%Matching the precise prescription behind the scale-adapted
%tables, and establishing an infinite-past limit, remain separate tasks.

The physical justification of the remote prescription differs from the
usual far-past construction in standard LQC.  In the latter case, moving the
initial surface sufficiently far into the low-curvature contracting phase
can place the relevant modes in an ultraviolet adiabatic regime, whereas at
or near the bounce adiabaticity need not select a unique state
\cite{Agullo:2017eyh}.

That last argument does not carry over to mLQC-I.  Its remote contracting
branch approaches de Sitter spacetime,
\begin{align}
 H&\longrightarrow-H_\Lambda,
 \quad
 a(t)\propto e^{-H_\Lambda t},
 \nonumber\\
 \eta&\simeq\frac{1}{aH_\Lambda}\longrightarrow0^+,
 \quad t\to-\infty.
 \label{eq:remote-dS-asymptotics}
\end{align}
Consequently, for every fixed finite comoving wave number,
$k/(a|H|)\simeq k\eta\to0$ rather than infinity.  The remote de Sitter state
used here is therefore \emph{not} justified by a far-past adiabatic
short-wavelength limit.  Rather, the asymptotic de Sitter structure supplies
the relevant state-selection principle \cite{Pan:2025ppa}.

\subsubsection{Numerical convergence and initial-time stability}

The finite-start tests at $t_i=-10,-15,-20,-25\,t_{\rm P}$ are an
initial investigation of the remote prescription. They should not be
interpreted as a far-past adiabatic limit: every fixed finite comoving mode
becomes nonadiabatic as $t\to-\infty$.

We supplement them with the independent calculation in
Appendix~\ref{app:independent-validation}. It fixes
$\epsilon_\Lambda=4.085090207021769\times10^{-5}$ and evaluates the full
signed convolution at $k=0.3m_{\rm P}$ on the smaller control window
$0.05\leq p/m_{\rm P},q/m_{\rm P}\leq1.05$, with triangle inequalities
and $N_f=18$. The critical post-bounce time integral is evaluated by
validated local phase-aware quadrature. The retained result is
$\Xi\simeq3.45\times10^{-4}$. On matched momentum grids, fresh starts
at $t_i/t_{\rm P}=-20,-25,-30$ change this estimate by at most approximately
$0.012\%$ relative to the $-25$ reference. All sixteen critical time panels
and their independent final-sample tests pass the stated numerical gates.

% This is conditional evidence for the retained correction in that fixed
% window, not a replacement for the scale-adapted values below. Each fresh
% start changes both the approximate free state and the zero connected
% three-point initial condition. A separate fixed-free-state interaction-cutoff
% test is therefore reported in Appendix~\ref{app:validation-remote-result}.
% Neither comparison establishes a unique infinite-past interacting state;
% deep fresh initialization of a frozen asymptotic formula remains sensitive
% to subleading matching data. The numerical indicators are not certified
% total error bounds and do not control all omitted same-order terms.

\subsubsection{Scale-adapted comparison with the bounce-defined state}
The remote-state convolution is substantially more oscillatory than the
bounce-state one.  To compare the two prescriptions without introducing a
fixed momentum interval that has very different meaning for different
external scales, we use the scale-adapted numerical control window
\begin{equation}
 0.01\,k\leq p,q\leq3k.
 \label{eq:selfsimilar-window}
\end{equation}
The endpoints in Eq.~(\ref{eq:selfsimilar-window}) remain numerical controls,
not physical cutoffs.  Table~\ref{tab:selfsimilar-Xi} shows the magnitude of the signed
correction for both initial states.  The bounce-defined result is extremely
small, whereas the remote result decreases rapidly as the external momentum
moves toward the bounce-curvature scale at the sampled momenta. Since the
integration window itself varies with $k$, this trend describes the specified
moving-window diagnostic rather than an unrestricted integral.
For reference we define
\begin{equation}
 k_B\equiv a_B\sqrt{\frac{R_B}{6}}
 \simeq1.60094\,m_{\rm P}.
 \label{eq:kB-num}
\end{equation}

%as the original adopted scale used for the table centers.
\begin{table}[!tbp]

 \caption{Magnitude of the signed correlation-function correction on the common scale-adapted numerical window
 in Eq.~(\ref{eq:selfsimilar-window}).  The external momentum is expressed in
 units of the bounce-curvature scale $k_B$.}
 \label{tab:selfsimilar-Xi}
 \begin{ruledtabular}
 \begin{tabular}{ccc}
 $k/k_B$ & $\Xi_{\rm bounce}$ & $\Xi_{\rm remote}$ \\
 \hline
 0.10 & $4.80\times10^{-9}$ & $9.88\times10^{-2}$ \\
 0.30 & $7.54\times10^{-9}$ & $7.87\times10^{-3}$ \\
 0.50 & $9.01\times10^{-10}$ & $1.40\times10^{-3}$ \\
 1.00 & $8.51\times10^{-9}$ & $2.30\times10^{-4}$
 \end{tabular}
 \end{ruledtabular}
\end{table}

The remote signed result is cancellation dominated.  A useful way of
quantifying the sensitivity to the rapidly varying external phase is to
average the correction $\delta\equiv\delta_B+\delta_W$ with the leading power spectrum as weight,
\begin{equation}
 \bar\delta(k)
 \equiv
 \frac{\langle\calP_{\calR}^{(0)}\,\delta\rangle_{\rm phase}}
 {\langle\calP_{\calR}^{(0)}\rangle_{\rm phase}},
 \quad
 \bar\Xi\equiv|\bar\delta|.
 \label{eq:phase-avg-Xi}
\end{equation}

% \begingroup\color{blue}\label{rv38}
% Write $\bar\delta=\langle\delta\rangle_w$ for the normalized positive
% power weight in Eq.~(\ref{eq:phase-avg-Xi}). Then
% \[
%  |\langle\delta\rangle_w|\leq\langle|\delta|\rangle_w
%  \leq\sup_{\rm interval}|\delta|.
% \]
%  A small signed average bounds neither the absolute mean nor the
% maximum. For samples, $M_n=\max_j|\delta(\theta_j)|$ is the maximum sampled,
% not a continuous maximum. 
% \par\endgroup
 For the remote state we find
\begin{align}
 \bar\Xi={}&5.05\times10^{-2},\ 6.82\times10^{-3},\ 8.88\times10^{-4},
 \nonumber\\
 &9.72\times10^{-5},\ 1.76\times10^{-5},
 \label{eq:phase-avg-values}
\end{align}
for $k/k_B=0.1,0.3,0.5,0.7,1$, respectively.  The absolute-integrand
quantity $\Xi_{\rm abs}$ is not small for the remote prescription, so the
small signed result cannot be promoted to a cancellation-independent bound.
The small signed averages do not bound pointwise corrections, and numerical
convergence of the cancellation-dominated integral requires further tests.
The reported trend refers to the stated scale-adapted window and sampled momenta.

\label{v5removed:26}

\section{Primordial non-Gaussianity and CMB projection}
\renewcommand{\theequation}{6.\arabic{equation}}
\setcounter{equation}{0}
\label{sec:ng-projection}

\subsection{Primordial power spectra}
\label{subsec:power-spectra}

Before discussing the three-point function, it is useful to display the
late-time leading-order curvature power spectrum for the two initial-state
prescriptions.  Because the spectra contain rapid local oscillations, the figures report a local phase average and phase standard deviation. These averages are used
only to display the slowly varying envelope; the correlation-function
convolutions themselves use the resolved mode functions.

\begin{figure*}[!tbp]
 \centering
 \includegraphics[width=0.48\textwidth]{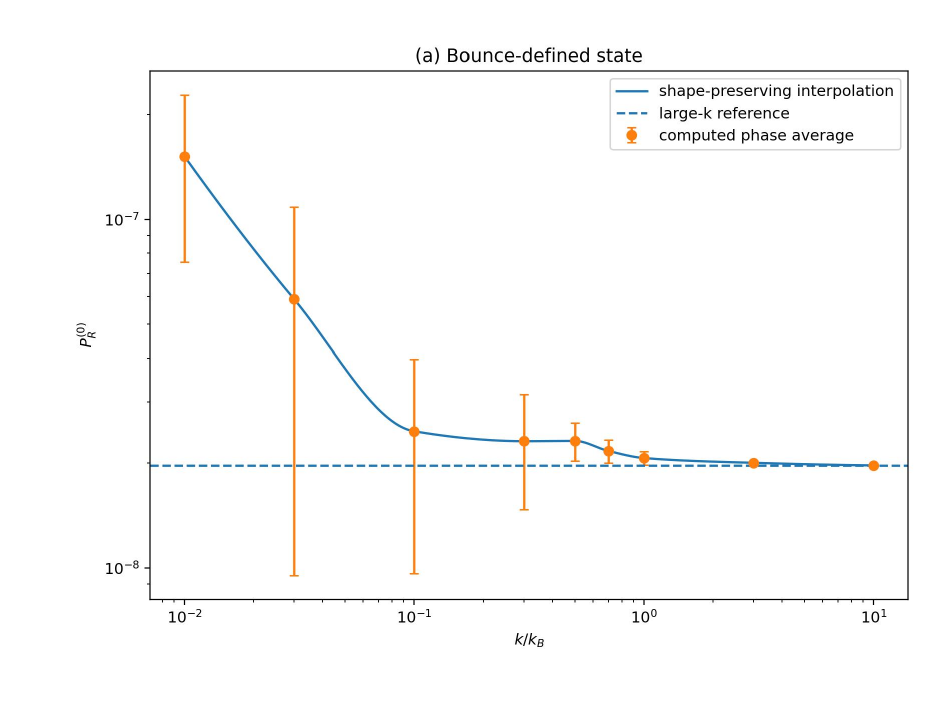}\hfill
 \includegraphics[width=0.48\textwidth]{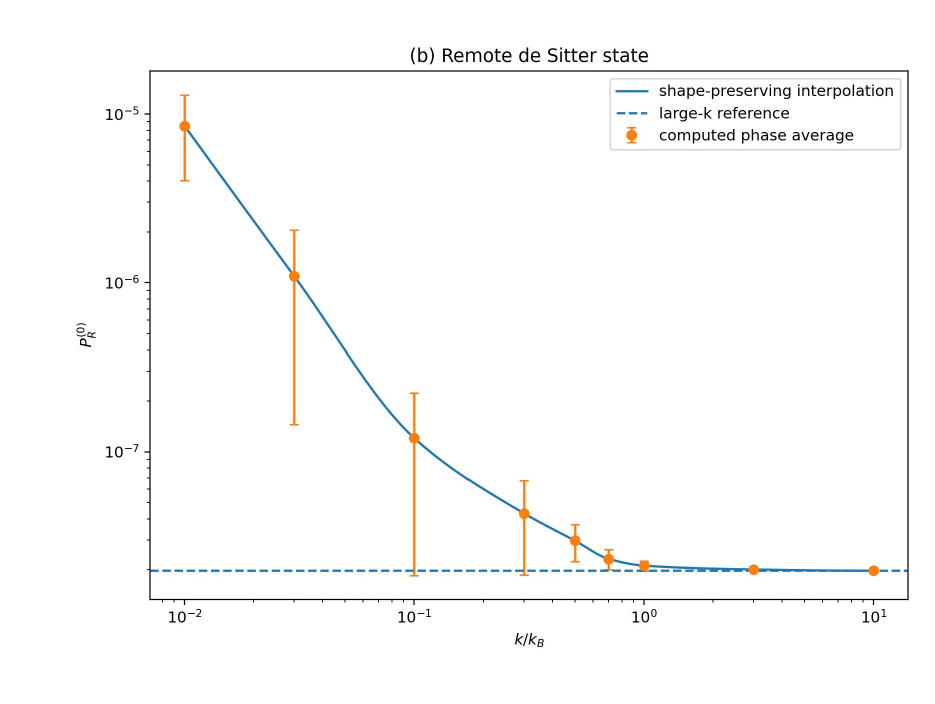}
 \caption{Late-time leading-order curvature power spectrum for
 (a) the bounce-defined state and (b) the remote de Sitter state.  
 Markers show the phase-averaged $\mathcal P_{\cal R}^{(0)}$ at the
 sampled values of $k/k_B$, error bars show one phase standard deviation, and
 the smooth curve is a shape-preserving interpolation through those computed
 points.  The dashed line is the large-$k$ reference set by the last sampled
 point.  The strong infrared
 enhancement, particularly for the remote state, is a primordial
 momentum-space feature and should not by itself be identified with an
 observable CMB enhancement.}
 \label{fig:power-two-states}
\end{figure*}

\subsection{Curvature bispectrum and non-Gaussianity parameter}
\label{subsec:fnl-def}

For the later analysis of the primordial three-point function it is useful to
state explicitly the convention used for $f_{\rm NL}$.  We define
\begin{equation}
 P_q(k)\equiv \left|\frac{u_k}{z_s}\right|^2,
\end{equation}
and the symmetric quadratic combination
\begin{align}
 \Sigma_q(k_1,k_2,k_3)
 \equiv{}&P_q(k_1)P_q(k_2)+P_q(k_1)P_q(k_3)
 \nonumber\\
 &+P_q(k_2)P_q(k_3).
 \label{eq:Sigma-q}
\end{align}
Equation~(\ref{eq:R-Fourier}) then gives, at tree level,
\begin{equation}
 B_{\calR}(k_1,k_2,k_3)
 =-
 \frac{B_\nu(k_1,k_2,k_3)}{z_s^3}
 +2C_{\calR}\Sigma_q(k_1,k_2,k_3).
 \label{eq:BR-tree}
\end{equation}
The overall minus sign in the intrinsic term follows from the linear
convention $\calR^{(1)}=-\nu/z_s$ established in
Eq.~(\ref{eq:R-linear-classical}).  We define
\begin{equation}
 f_{\rm NL}(k_1,k_2,k_3)
 =-
 \frac{5}{6}
 \frac{B_{\calR}(k_1,k_2,k_3)}{\Sigma_q(k_1,k_2,k_3)}.
 \label{eq:fNL-def}
\end{equation}
For equilateral configurations this reduces to
\begin{equation}
 f_{\rm NL}^{\rm eq}(k)
 =-
 \frac{5}{18}
 \frac{B_{\calR}(k,k,k)}{P_q(k)^2}.
 \label{eq:fNL-eq}
\end{equation}
The nonlinear-conversion piece alone contributes
$f_{\rm NL}^{\rm conv}=-(5/3)C_{\calR}$, which is of slow-roll order on the
late evaluation surface.

\subsection{Large primordial non-Gaussianity}
\label{subsec:primordial-fNL}

\begin{figure*}[!tbp]
 \centering
 \includegraphics[width=0.78\textwidth]{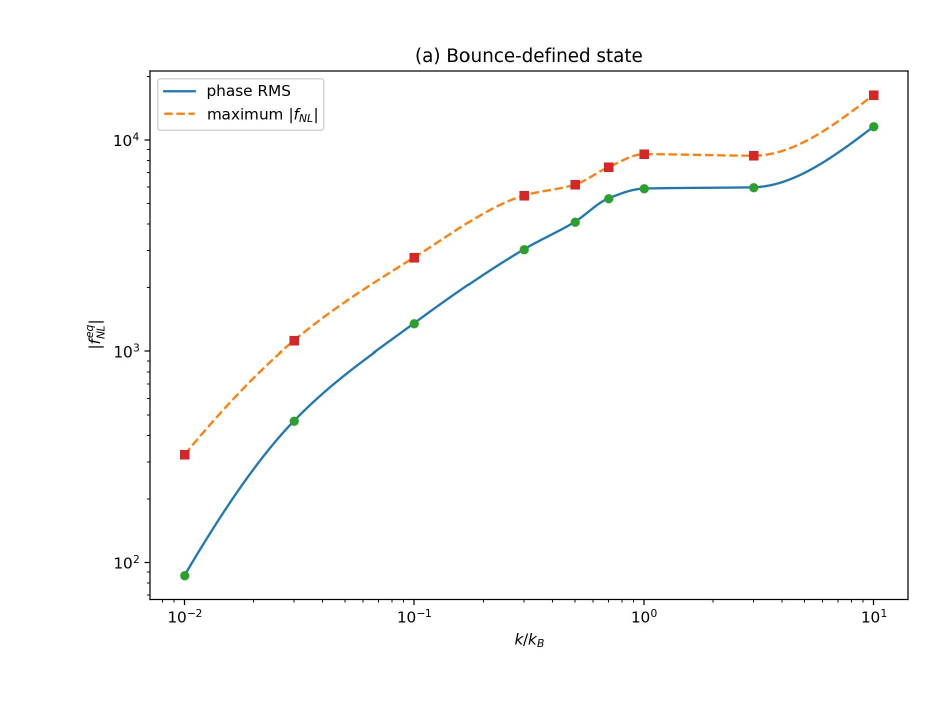}
 \includegraphics[width=0.78\textwidth]{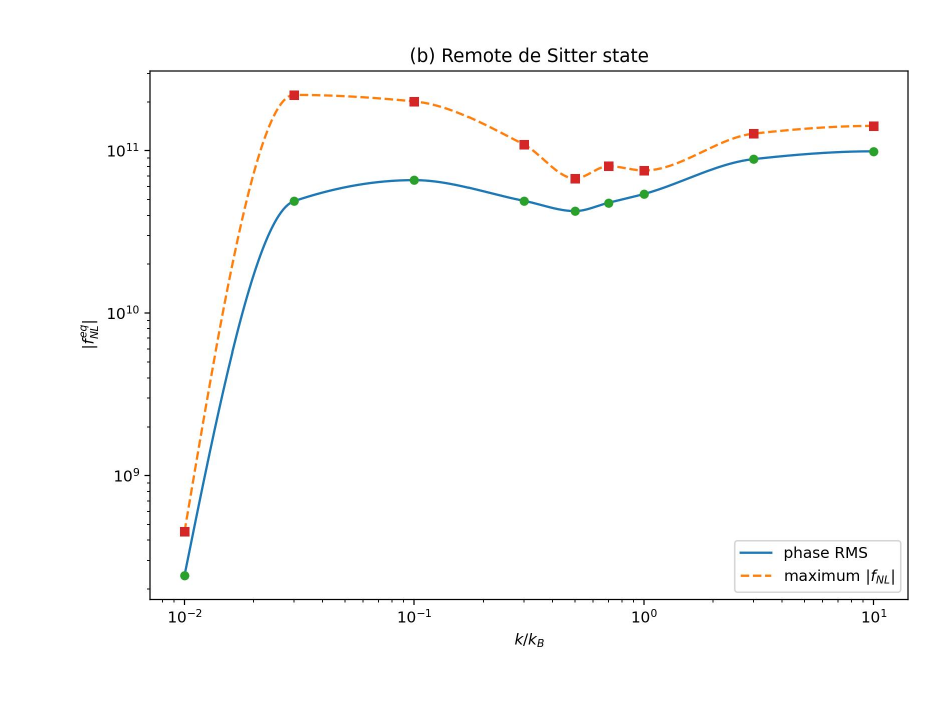}
 \caption{Equilateral primordial non-Gaussianity for (a) the bounce-defined state and (b) the remote de Sitter state. The markers report local RMS amplitude and maximum sampled $|f_{\rm NL}^{\rm eq}|$. 
 The solid and dashed curves are shape-preserving interpolants through those points. These are primordial momentum-space quantities, not observed CMB amplitudes.}
 \label{fig:fNL-envelope}
\end{figure*}

The small retained correction to the power spectrum does not imply a small
pointwise three-point function.  Figure~\ref{fig:power-two-states} provides the
corresponding two-point baseline for both initial-state prescriptions.  Using Eqs.~(\ref{eq:BR-tree})--
(\ref{eq:fNL-eq}), the equilateral bispectrum is highly oscillatory.  To characterize its envelope we retain the originally reported local RMS
values of $f_{\rm NL}^{\rm eq}$ in Table~\ref{tab:fNL-envelope}.  The bounce-defined state gives values of order
$10^3$--$10^4$ over the displayed range, whereas the remote state gives much
larger pointwise values.  These numbers should not be interpreted as CMB
constraints because the oscillations are much faster than the relevant
projection scales.  The corresponding equilateral envelopes are shown in
Fig.~\ref{fig:fNL-envelope}; the figure makes the large separation between the
two state prescriptions, as well as the oscillatory-envelope character of the
quoted amplitudes, directly visible.

\begin{table}[!tbp]

 \caption{Phase-RMS equilateral primordial non-Gaussianity.  The
 entries characterize the primordial pointwise envelope, not an observed CMB amplitude. }
 \label{tab:fNL-envelope}
 \begin{ruledtabular}
 \begin{tabular}{ccc}
 $k/k_B$ & $f_{\rm NL,RMS}^{\rm bounce}$ & $f_{\rm NL,RMS}^{\rm remote}$ \\
 \hline
 0.10 & $1.35\times10^{3}$ & $6.59\times10^{10}$ \\
 0.30 & $3.04\times10^{3}$ & $4.91\times10^{10}$ \\
 1.00 & $5.89\times10^{3}$ & $5.41\times10^{10}$ \\
 3.00 & $5.97\times10^{3}$ & $8.88\times10^{10}$ \\
 10.0 & $1.16\times10^{4}$ & $9.92\times10^{10}$
 \end{tabular}
 \end{ruledtabular}
\end{table}

\subsection{Rapid \texorpdfstring{$k_t$}{kt} phase and momentum-triangle dependence}
A dense scan of the raw curvature bispectrum shows that the dominant rapid
phase is organized by the total momentum
\begin{equation}
 k_t\equiv k_1+k_2+k_3.
\end{equation}
Across equilateral, squeezed, flattened, and generic triangles, the measured
period is
\begin{equation}
 \Delta k_t\simeq5.7\times10^{-4}m_{\rm P},
 \quad
 k_I^{\rm eff}\equiv\frac{\Delta k_t}{2\pi}
 \simeq9.1\times10^{-5}m_{\rm P}.
 \label{eq:kI-eff}
\end{equation}
A convenient numerical representation is therefore
\begin{align}
 f_{\rm NL}(k_1,k_2,k_3)
 \simeq{}&
 F_{\rm eq}(k_t)\,
 S_f(k_t;r_1,r_2,r_3)
 \nonumber\\
 &\times
 \sin\left[
 \frac{k_t}{k_I^{\rm eff}}
 +\varphi(k_t;r_1,r_2,r_3)
 \right],
 \label{eq:mLQCI-template}
\end{align}
where $r_i=k_i/k_t$ and $r_1+r_2+r_3=1$; a state label on
$F_{\rm eq}$, $S_f$, and $\varphi$ is suppressed in this equation.  The
rapidly varying part is controlled by $k_t$, whereas these three functions
vary slowly
on the scale $\Delta k_t$.  We verified this structure on a grid of 37
triangle shapes at $k_t=3,8,16,$ and $24\,m_{\rm P}$.  For the bounce-defined
state the median variation of the normalized shape amplitude over this scale
range is about $16\%$, and the median variation of the relative phase is
about $0.19$ rad.  The corresponding remote-state values are approximately
$24\%$ and $0.20$ rad, with larger excursions confined mainly to the most
squeezed configurations.

\label{v5removed:32}

\subsection{General fitting functions for the oscillatory template}
\label{subsec:template-fit}

\begin{figure*}[!tbp]
 \centering
 \includegraphics[width=0.78\textwidth]{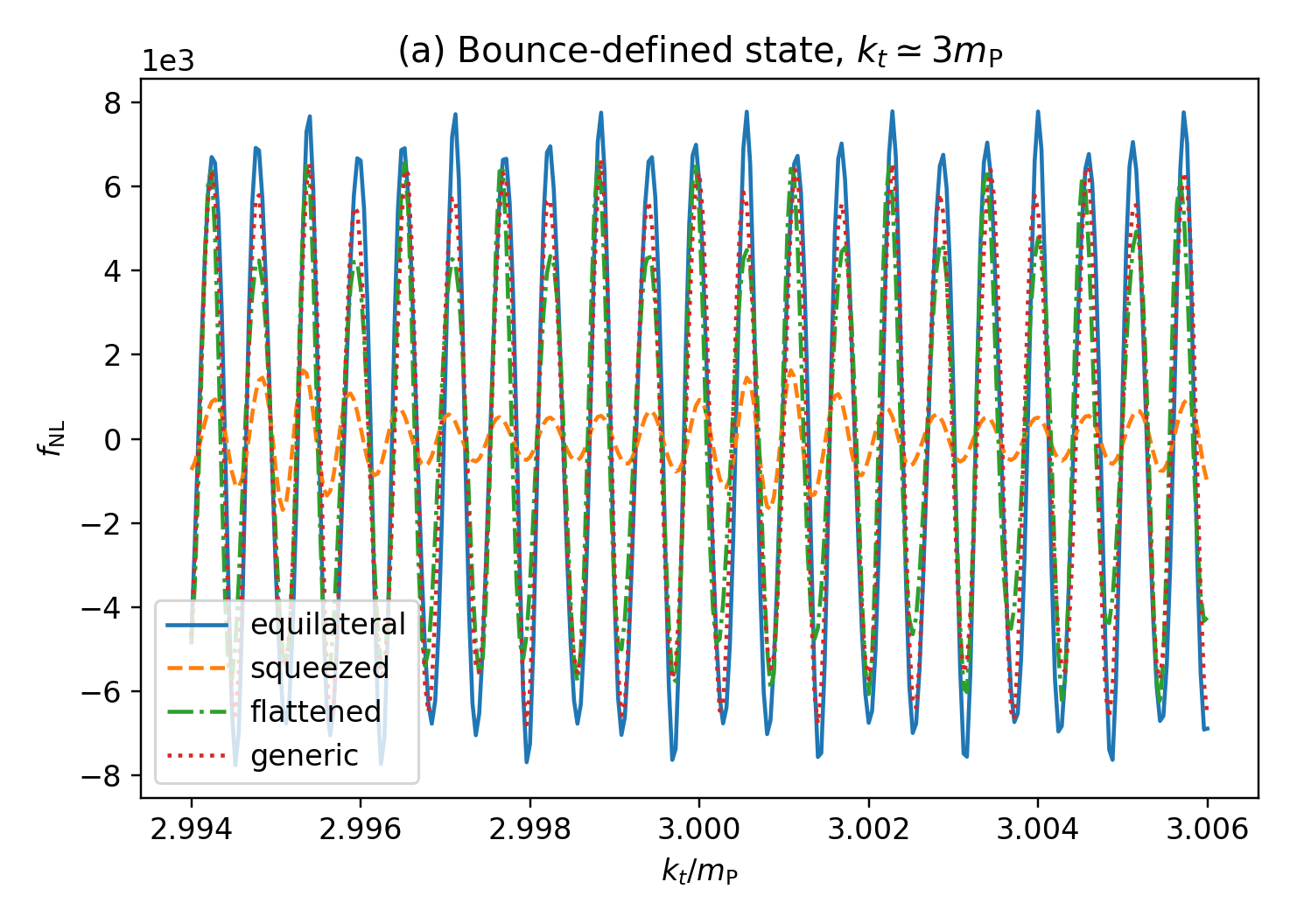}
 \includegraphics[width=0.78\textwidth]{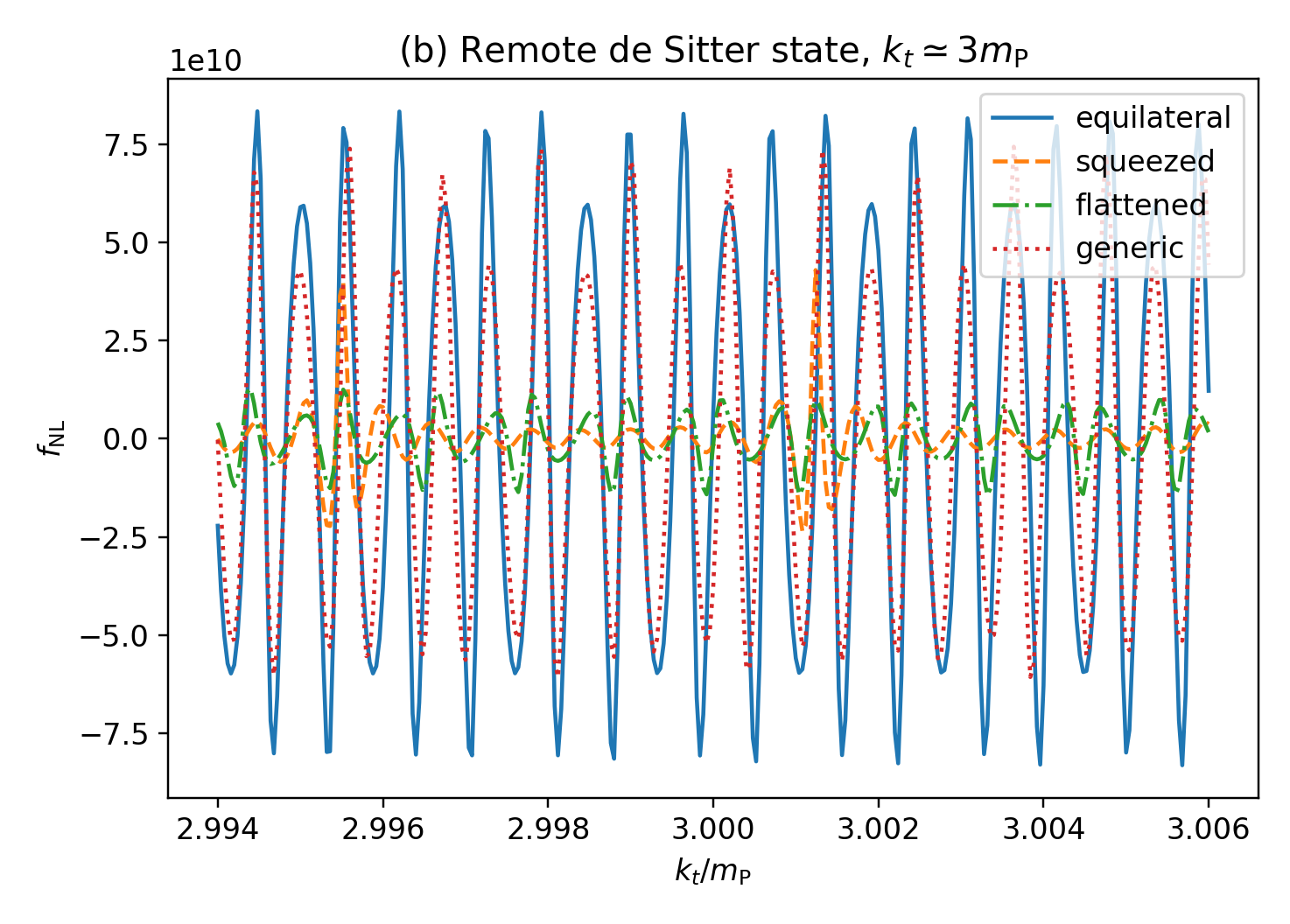}
 \caption{Resolved primordial $f_{\rm NL}$ oscillations in a narrow interval
 around $k_t=3m_{\rm P}$ for (a) the bounce-defined state and (b) the remote
 de Sitter state.  The four configurations are
 equilateral $(1/3,1/3,1/3)$, squeezed $(0.475,0.475,0.05)$,
 flattened $(0.49,0.255,0.255)$, and generic $(0.40,0.35,0.25)$, where the
 triplets denote $(k_1,k_2,k_3)/k_t$.  The nearly common rapid period provides
 direct numerical evidence that the leading oscillatory phase is controlled
 primarily by $k_t=k_1+k_2+k_3$, while the shape modulates the amplitude and
 subleading phase.}
 \label{fig:fNL-shapes-kt3}
\end{figure*}

Equation~(\ref{eq:mLQCI-template}) is intended to describe the general
momentum-triangle dependence over the sampled domain, not only the four
representative configurations shown in Fig.~\ref{fig:fNL-shapes-kt3}.  The
latter are used only to display the oscillations visually.  The numerical fit
uses 37 triangle shapes at each of the four reference values
$k_t=3,8,16,$ and $24\,m_{\rm P}$ for each initial-state prescription, giving
$N_{\rm node}=148$ nodes for each state.  The current shape grid reaches
$r_{\min}=0.05$; more strongly squeezed triangles are outside the presently
validated interpolation domain.

For the slowly varying equilateral envelope we introduce
\begin{equation}
 x\equiv\ln\left(\frac{k_t}{8m_{\rm P}}\right)
\end{equation}
and use the compact cubic representation
\begin{equation}
 F_{\rm eq}^{(s)}(k_t)
 =
 \exp\left[
 a_0^{(s)}+a_1^{(s)}x+a_2^{(s)}x^2+a_3^{(s)}x^3
 \right],
 \label{eq:Feq-fit}
\end{equation}
where $s=B,R$ denotes the bounce-defined and remote de Sitter states,
respectively.  The coefficients obtained from the four equilateral reference
scales are listed in Table~\ref{tab:Feq-fit}.
\begin{table}[!tbp]
\caption{Coefficients in the equilateral-envelope fit
(\ref{eq:Feq-fit}).}
\label{tab:Feq-fit}
\begin{ruledtabular}
\begin{tabular}{ccccc}
state & $a_0$ & $a_1$ & $a_2$ & $a_3$\\
\hline
bounce & $9.05327$ & $-0.06662$ & $-0.02604$ & $0.23276$\\
remote & $25.22567$ & $0.53570$ & $0.14966$ & $-0.09670$
\end{tabular}
\end{ruledtabular}
\end{table}
The cubic form interpolates the four quoted equilateral amplitudes exactly
on $3\leq k_t/m_{\rm P}\leq24$; it should not be extrapolated outside this
interval.  Here $F_{\rm eq}$ denotes the fitted sinusoidal amplitude entering
Eq.~(\ref{eq:mLQCI-template}) and should not be confused with the phase-RMS
equilateral amplitude quoted in Table~\ref{tab:fNL-envelope}.  Additional
$k_t$ nodes are required before interpreting Eq.~(\ref{eq:Feq-fit}) as a
sub-percent interpolation between the reference scales.

For the general triangle dependence it is useful to impose permutation
symmetry explicitly.  We therefore introduce the two symmetric shape
coordinates
\begin{equation}
 U\equiv
 6\sum_{i=1}^3\left(r_i-\frac13\right)^2,
 \quad
 V\equiv
 27\prod_{i=1}^3\left(r_i-\frac13\right),
 \label{eq:UV-shape}
\end{equation}
which both vanish for the equilateral triangle.  To organize the amplitude
and phase without artificial $2\pi$ branch jumps, we define the complex slow
envelope
\begin{equation}
 {\cal G}^{(s)}(k_t;U,V)
 \equiv
 S_f^{(s)}(k_t;U,V)
 \exp\!\left[i\Delta\varphi^{(s)}(k_t;U,V)\right],
 \label{eq:Gshape}
\end{equation}
with
${\cal G}^{(s)}(k_t;0,0)=1$ and a continuously unwrapped relative phase.
The full slow phase in Eq.~(\ref{eq:mLQCI-template}) is decomposed as
\begin{align}
 \varphi^{(s)}(k_t;U,V)
 &=
 \varphi_{\rm eq}^{(s)}(k_t)
 +\Delta\varphi^{(s)}(k_t;U,V),
 \nonumber\\
 \Delta\varphi^{(s)}(k_t;0,0)&=0.
 \label{eq:phase-decomposition}
\end{align}
The dominant equilateral phase is already represented by
$k_t/k_I^{\rm eff}$ in Eq.~(\ref{eq:mLQCI-template}); the residual
$\varphi_{\rm eq}^{(s)}(k_t)$ varies slowly but is evaluated numerically in
the present analysis rather than assigned a global polynomial fit from only
four $k_t$ reference nodes.

For numerical work we first interpolate two unconstrained smooth functions,
$\widetilde A^{(s)}$ and $\widetilde\Phi^{(s)}$, in the three variables
$(x,U,V)$ using a permutation-invariant thin-plate radial-basis
representation,
\begin{align}
 \widetilde A^{(s)}({\bf y})
 &=
 p_A^{(s)}({\bf y})
 +\sum_{j=1}^{N_{\rm node}}
 w_{Aj}^{(s)}
 \Phi\!\left(\left|{\bf y}-{\bf y}_j\right|\right),
 \nonumber\\
 \widetilde\Phi^{(s)}({\bf y})
 &=
 p_\varphi^{(s)}({\bf y})
 +\sum_{j=1}^{N_{\rm node}}
 w_{\varphi j}^{(s)}
 \Phi\!\left(\left|{\bf y}-{\bf y}_j\right|\right),
 \label{eq:RBF-template}
\end{align}
where
\begin{equation}
 {\bf y}=(x,U,V),
 \quad
 \Phi(\rho)=\rho^2\ln\rho,
 \quad
 \Phi(0)=0.
 \label{eq:RBF-kernel}
\end{equation}
The polynomial completion is taken to be linear in $(x,U,V)$,
\begin{equation}
 p_X^{(s)}({\bf y})
 =
 c_{X0}^{(s)}
 +c_{Xx}^{(s)}x
 +c_{XU}^{(s)}U
 +c_{XV}^{(s)}V,
 \quad
 X=A,\varphi ,
 \label{eq:RBF-poly}
\end{equation}
and the corresponding coefficients are given explicitly in
Appendix~\ref{app:RBF-complete}.

The radial weights are determined from the standard thin-plate system
\begin{equation}
 \begin{pmatrix}
 K+\lambda_X^{(s)} I & P\\
 P^T & 0
 \end{pmatrix}
 \begin{pmatrix}
 {\bf w}_X^{(s)}\\
 {\bf c}_X^{(s)}
 \end{pmatrix}
 =
 \begin{pmatrix}
 {\bf d}_X^{(s)}\\
 0
 \end{pmatrix},
 \label{eq:RBF-linear-system}
\end{equation}
where
\begin{equation}
 K_{ij}
 =
 \Phi\!\left(|{\bf y}_i-{\bf y}_j|\right),
 \quad
 P_j=(1,x_j,U_j,V_j).
\end{equation}
For the amplitude fit,
$d_{A,j}=\ln S_{f,j}$, while for the phase fit
$d_{\varphi,j}=\Delta\varphi_j$, with both quantities defined relative to
the equilateral configuration at the same $k_t$.  The smoothing parameters
used in the fits are
\begin{align}
 \lambda_A^{(B)}&=10^{-5},
 &\lambda_\varphi^{(B)}&=10^{-5},
 \nonumber\\
 \lambda_A^{(R)}&=10^{-5},
 &\lambda_\varphi^{(R)}&=10^{-4}.
 \label{eq:RBF-smoothing}
\end{align}
Appendix~\ref{app:RBF-complete} summarizes the interpolation prescription.
% and gives the polynomial coefficients used in the numerical analysis.  The printed
% polynomial coefficients alone do not reconstruct the interpolation: the node
% coordinates, radial weights or target data, and equilateral residual phase
% are also required.

We then impose the equilateral normalization exactly at every $k_t$ by
defining
\begin{align}
 \ln S_f^{(s)}(x,U,V)
 &=
 \widetilde A^{(s)}(x,U,V)
 -\widetilde A^{(s)}(x,0,0),
 \nonumber\\
 \Delta\varphi^{(s)}(x,U,V)
 &=
 \widetilde\Phi^{(s)}(x,U,V)
 -\widetilde\Phi^{(s)}(x,0,0).
 \label{eq:RBF-normalization}
\end{align}
Thus, we have
\begin{equation}
 S_f^{(s)}(x,0,0)=1,
 \quad
 \Delta\varphi^{(s)}(x,0,0)=0
\end{equation}
identically, rather than only at the sampled equilateral nodes.  The
representation therefore treats the four configurations in
Fig.~\ref{fig:fNL-shapes-kt3} as ordinary points in a single
permutation-invariant fit over the sampled triangle domain.

The 37-shape grid motivates the representation in
Eq.~(\ref{eq:mLQCI-template}), but a common oscillation period does not by
itself establish the accuracy of the single-harmonic waveform. The grid does
not yet justify a uniform
sub-percent interpolation error over the full sampled domain.  Applying a
leave-one-node-out test to the normalized RBF prescription above gives
median errors
\begin{equation}
 {\rm median}\,|\delta\ln S_f|
 \simeq
 5.1\times10^{-3}\;(B),
 \;\;
 3.64\times10^{-2}\;(R),
 \label{eq:RBF-CV-amp}
\end{equation}
with corresponding 90th-percentile values
$3.70\times10^{-2}$ and $1.44\times10^{-1}$.  For the relative phase,
\begin{equation}
 {\rm median}\,|\delta\Delta\varphi|
 \simeq
 6.5\times10^{-3}\ {\rm rad}\;(B),
 \;\;
 3.3\times10^{-3}\ {\rm rad}\;(R),
 \label{eq:RBF-CV-phase}
\end{equation}
while the corresponding 90th-percentile errors are approximately
$0.036$ and $0.071$ rad.  The largest interpolation errors remain
concentrated toward the squeezed boundary.  We therefore distinguish the
accuracy of the physical calculation at the directly computed nodes from
the interpolation accuracy between them.  A genuinely uniform sub-percent
template will require an adaptively refined shape and $k_t$ grid, with the
squeezed region sampled more densely.

\subsection{CMB-projection diagnostics}
\label{subsec:CMB-projection}

The observed reduced CMB bispectrum involves the radiation-transfer
functions and spherical-Bessel kernels.  For temperature anisotropies, we have
\begin{align}
 b^{TTT}_{\ell_1\ell_2\ell_3}
 ={}&
 \left(\frac{2}{\pi}\right)^3
 \int x^2\dd x
 \prod_{i=1}^3\left[\int k_i^2\dd k_i\right]
 B_{\calR}(k_1,k_2,k_3)
 \nonumber\\
 &\times
 \prod_{i=1}^3
 \left[
 \Delta^T_{\ell_i}(k_i)j_{\ell_i}(k_i x)
 \right].
 \label{eq:CMB-reduced-bispectrum}
\end{align}
Rapid primordial oscillations can be strongly averaged by the CMB
projection, whereas smoother bouncing templates can remain observable after
transfer projection \cite{Roshna:2023,vanTent:2023}.  The momentum dependence
of the primordial bispectrum must therefore be retained rather than reduced
to a single pointwise amplitude.

A full CLASS transfer-function calculation is beyond the scope of the
present paper.  We instead consider two approximate projection diagnostics.
First, in a thin-last-scattering-shell approximation we retain the
spherical-Bessel kernels and the full numerical shape interpolation
$F_{\rm eq}S_f e^{i\varphi}$.  Second, we replace the purely geometric kernel
by the acoustic model
\begin{align}
 \Delta^T_\ell(k)
 \propto{}&
 e^{-(k/k_D)^2}
 e^{-k^2\sigma_*^2/2}
 \Big[
 \cos(kr_s)j_\ell(kD_*)
 \nonumber\\
 &+\beta\sin(kr_s)j_\ell'(kD_*)
 \Big],
 \label{eq:acoustic-transfer-model}
\end{align}
with
$D_*=1.4\times10^4\,{\rm Mpc}$,
$r_s=144.4\,{\rm Mpc}$,
$k_D=0.14\,{\rm Mpc}^{-1}$,
$\sigma_*=15\,{\rm Mpc}$, and baseline $\beta=0.5$.
This model is intended only to test whether ordinary acoustic and damping
scales spoil the rapid-phase cancellation; it is not a substitute for a
Boltzmann code.

For the oscillatory template it is useful to normalize the projected
residual to a phase-unsuppressed absolute-weight reference,
\begin{equation}
 {\cal S}_{\ell_1\ell_2\ell_3}
 \equiv
 \frac{|b^{\rm osc}_{\ell_1\ell_2\ell_3}|}
 {\int \dd\Pi\,|W_{\ell_1\ell_2\ell_3}B_{\rm env}|},
 \label{eq:projection-suppression}
\end{equation}
where $\dd\Pi$ denotes the numerical momentum measure, $W$ contains the
projection kernels, and $B_{\rm env}$ is the adopted reference envelope.
% \textcolor{blue}{Removing the sine from $f_{\rm NL}$ does not by itself define a slowly varying
% envelope of $B_{\cal R}=-(6/5)\Sigma_qf_{\rm NL}$ because $\Sigma_q$ also
% oscillates.  Eq.~(\ref{eq:projection-suppression})
% is retained as the definition of a reported oscillatory-only diagnostic.}  
This normalization avoids artificially large
ratios when the approximate acoustic transfer function changes sign.

\begin{figure}[!tbp]
 \centering
 \includegraphics[width=0.45\textwidth]{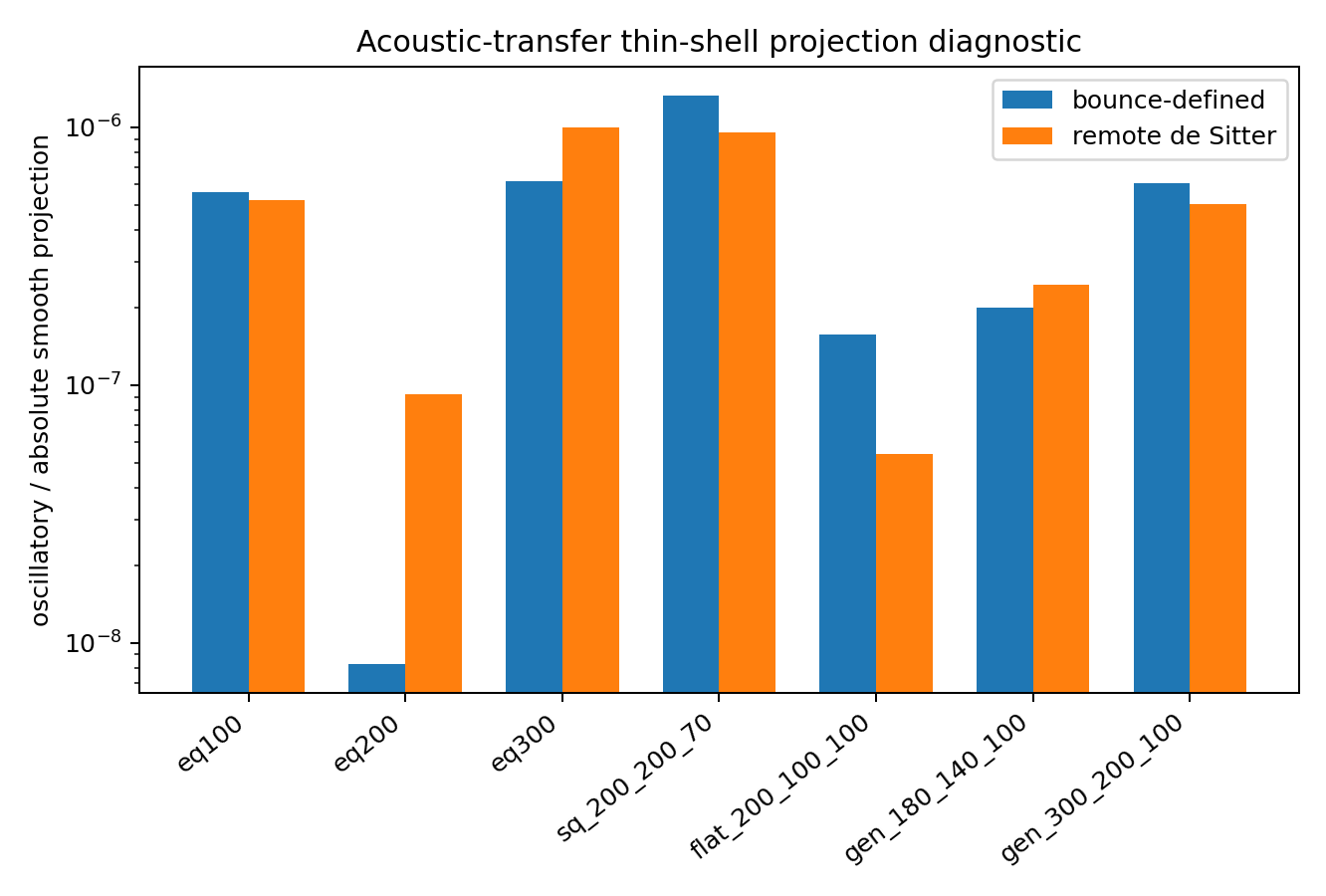}
 \caption{Baseline acoustic thin-shell projection diagnostic for several
 representative multipole configurations.  The normalization uses the
 absolute phase-unsuppressed weight in
 Eq.~(\ref{eq:projection-suppression}); hence the figure measures the
 cancellation associated with the rapid primordial phase rather than a ratio
 to a particular standard CMB template.}
 \label{fig:acoustic-projection}
\end{figure}

For an illustrative bounce-to-present conversion, we take
$1\,m_{\rm P}$ at the bounce to correspond to approximately
$5\times10^{-3}\,{\rm Mpc}^{-1}$ today.  With this adopted mapping,
Eq.~(\ref{eq:kI-eff}) gives
\begin{equation}
 k_I^{\rm eff}\simeq4.57\times10^{-7}\,{\rm Mpc}^{-1},
 \quad
 \Delta k_t\simeq2.87\times10^{-6}\,{\rm Mpc}^{-1}.
 \label{eq:kI-today}
\end{equation}
For this illustrative mapping, the corresponding angular period is much smaller than one multipole:
$D_*\Delta k_t\simeq0.040$ when one momentum leg is varied with the others
fixed, and approximately $0.013$ along the common equilateral direction.
This is the ultra-rapid regime in which transfer projection can strongly
average the primordial phase.
%\begingroup\color{blue}\label{rv51}
For this illustrative mapping, Table~\ref{tab:projection-suppression} and
Fig.~\ref{fig:acoustic-projection} retain the reported projection outputs.

%\par\endgroup
\begin{table}[!tbp]
 \caption{Reported acoustic thin-shell projection diagnostic.  The quantity
 ${\cal S}$ is defined in Eq.~(\ref{eq:projection-suppression}) and is not an
 observed $f_{\rm NL}$.}
 \label{tab:projection-suppression}
 \begin{ruledtabular}
 \begin{tabular}{ccc}
 $(\ell_1,\ell_2,\ell_3)$ & ${\cal S}_{\rm bounce}$ & ${\cal S}_{\rm remote}$ \\
 \hline
 $(100,100,100)$ & $4.54\times10^{-7}$ & $4.56\times10^{-7}$ \\
 $(200,200,200)$ & $4.28\times10^{-7}$ & $4.66\times10^{-7}$ \\
 $(200,200,70)$  & $6.84\times10^{-7}$ & $4.49\times10^{-7}$
 \end{tabular}
 \end{ruledtabular}
\end{table}

% \begin{figure}[!tbp]

%  \centering
%  \includegraphics[width=\columnwidth]{mLQCI_Xi_abs_scan_enlarged.pdf}
%  \caption{Enlarged-window scan of the perturbativity diagnostics for
%  the bounce-defined state at $N=18$.  The conservative
%  $\Xi_{\rm abs}$ does not use cancellations in the oscillatory bispectrum
%  convolution and is estimated below $6\times10^{-6}$ at the five sampled
%  momenta.  The magnitude $\Xi$ of the signed correction is smaller but is more sensitive to momentum resolution. The legend\textquotesingle s ``signed $\Xi$\textquotesingle\textquotesingle\ denotes this magnitude.}
%  \label{fig:Xi-scan}
% \end{figure}

These illustrative diagnostics are consistent with substantial phase
averaging. The numerical residuals and their displayed precision require
independent convergence tests and propagation of primordial-template errors.  We deliberately
stop short of an observational constraint: a quantitative prediction requires
actual radiation-transfer functions, a full radial integration, and the same
primordial template evaluated over the complete transfer-supported momentum
domain.

\label{v5removed:39}

\section{Discussion and conclusions}
\renewcommand{\theequation}{7.\arabic{equation}}
\setcounter{equation}{0}
In this paper we have constructed and evaluated a retained
nonlinear-conversion diagnostic in hybrid mLQC-I.
  Starting from the classical
quadratic and cubic Hamiltonians in a common hybrid canonical representation,
we extended the sector-preserving mLQC-I prescription to the cubic theory and
constructed the exact finite-momentum interaction vertex.  The regular
Mukhanov--Sasaki variable $\nu$ is evolved through the quantum-gravity regime,
while the curvature perturbation is evaluated on a late super-Hubble surface
for the correlation-function calculation.  We have also retained the complete
classical second-order transformation between $\delta\phi$ and ${\cal R}$ and
given its minimal term-by-term mLQC-I effective continuation.  The latter is
not needed for the late-time numerical results reported here, but provides the
appropriate starting point for a future calculation of nonlinear correlation
functions in the contracting phase and through the bounce.

The first main result concerns the retained nonlinear-conversion correction to
the scalar two-point function.  For the representative bounce-defined initial state,
the cancellation-independent finite-window diagnostic satisfies
$\Xi_{\rm abs}<6\times10^{-6}$ at the five sampled external momenta.
%, subject to numerical convergence checks.
Thus, the retained nonlinear correction is very small even before making use
of oscillatory cancellations.  The remote de Sitter initial state is qualitatively
different.  Its signed correction is highly oscillatory and cancellation
dominated; on the common scale-adapted window it decreases from
$\Xi\simeq9.9\times10^{-2}$ at $k/k_B=0.1$ to
$\Xi\simeq2.3\times10^{-4}$ at $k/k_B=1$, and the phase-averaged diagnostic
shows the same rapid decrease.  Because $\Xi_{\rm abs}$ is not small for the
remote prescription, its perturbative control is less robust and cannot be
promoted to a cancellation-independent statement. 
% It should be noted that the signed values alone do not establish control of the full interacting
% correlation-function expansion; convergence of the cancellation-dominated
% integrals and the omitted interaction terms requires further study \cite{Saeed:2026xxx}.

An independent fixed-window calculation provides an additional numerical
check of the remote candidate: at $k=0.3m_{\rm P}$ and
$0.05\leq p/m_{\rm P},q/m_{\rm P}\leq1.05$, the retained result is
$\Xi\simeq3.45\times10^{-4}$. Local phase-aware time integration and
independent validation overcome the identified critical-interval time
aliasing in this conditional test. Its finite-start variation over
$t_i/t_{\rm P}=-20,-25,-30$ is at most about $0.012\%$ on matched grids;
see Appendix~\ref{app:independent-validation}. 
% The candidate state and
% fixed domain differ from the scale-adapted comparison, and the result
% neither replaces those entries nor establishes an infinite-past limit. A matched
% late-surface comparison gives a $1.48\%$ increase in the retained
% diagnostic between $N_F=18$ and $20$, closely tracking the conversion
% coefficient; it does not support strict late-time invariance
% [Appendix~\ref{app:validation-final-surface}].

% \begingroup\color{blue}
% The initial prescriptions have different motivations. The bounce-defined
% instantaneous quadratic ground state yields small reported finite-window
% conversion estimates; adiabaticity does not uniquely select a bounce state
% \cite{Agullo:2012fc,Agullo:2012sh,Agullo:2013ai}. The remote prescription is
% motivated by asymptotic de Sitter structure \cite{Pan:2025ppa}, but the hybrid
% equation, its finite-time matching and the relation to instantaneous
% diagonalization must be distinguished from the dressed-metric construction.
% For every fixed finite mode, $k/(a|H|)\to0$ as $t\to-\infty$; finite-start
% stability is not a far-past adiabatic-vacuum limit \cite{Saeed:2026vxw}.

% \par\endgroup

A second central result is that a small correction to the two-point function
can coexist with a very large pointwise primordial bispectrum.  In the
convention used here, the bounce-defined state has a phase-RMS equilateral
$f_{\rm NL}$ of order $10^3$--$10^4$ over the displayed range, while the
remote prescription can be many orders of magnitude larger.  This does not
contradict the correlation-function perturbativity test: the correction to the
two-point function contains an additional power of the small scalar power
spectrum, as summarized parametrically in
Eqs.~(\ref{eq:scaling-B}) and (\ref{eq:scaling-W}).
The reported finite-window calculations illustrate coexistence of a large
pointwise tree bispectrum and a small retained nonlinear-conversion
diagnostic. 
% Control of the complete interacting expansion requires the
% omitted same-order contributions and their numerical errors to be assessed.
% Conversely, large pointwise $f_{\rm NL}$ alone does not prove breakdown.

The momentum dependence of this large non-Gaussianity exhibits a simple and
useful structure.  Over the sampled triangle domain, the dominant rapid phase
is organized by the total momentum $k_t=k_1+k_2+k_3$, with
$\Delta k_t\simeq5.7\times10^{-4}m_{\rm P}$, whereas the triangle-shape
envelope and the residual phase vary much more slowly.  This motivates the
general representation in Eq.~(\ref{eq:mLQCI-template}).  We have formulated
a permutation-invariant fitting framework in terms of a smooth equilateral
envelope and a complex shape function.  

% The present 37-shape grid motivates this factorization, but the accuracy of
% the full single-harmonic waveform remains to be tested, and it does not yet
% provide a uniformly
% sub-percent interpolation over the sampled triangle domain: the
% leave-one-out amplitude errors are smallest for the bounce-defined state and
% increase toward the squeezed boundary, especially for the remote state.
% Accordingly, all primary primordial results quoted here are taken from the
% directly computed nodes.  A denser adaptive scan in both $k_t$ and triangle
% shape, with independent validation, will be required to turn this
% representation into a certified sub-percent primordial template.

For the illustrative bounce-to-present mapping adopted in
Sec.~\ref{subsec:CMB-projection}, the very short primordial phase scale has
an immediate consequence for CMB projection.  Thin-shell tests retaining the
numerical triangle dependence, followed by an acoustic transfer model
including monopole, Doppler, diffusion-damping, and finite-width effects, give
residual phase-cancellation factors of order $10^{-7}$--$10^{-6}$ for
representative configurations.
%\footnote{These values are illustrative: their numerical errors, template-domain dependence, and sensitivity to interpolation and boundary treatment remain tobe quantified.}. 
Rapid oscillations can be averaged by projection
\cite{Roshna:2023}, while smoother templates may survive more efficiently
\cite{vanTent:2023}.

The projection analysis is deliberately a diagnostic rather than an
observational prediction.  A definitive CMB constraint requires the exact
radiation-transfer functions, full radial integration, and a primordial
template validated throughout the complete transfer-supported momentum
domain \cite{Roshna:2023}.  

Likewise, the present correlation-function calculation does not
establish the absence of every possible strong-coupling or ultraviolet
consistency issue \cite{Saeed:2026xxx}.  Genuine ${\cal{O}}\left(\left[H_I^{(3)}\right]^2\right)$ loop corrections, quartic and
higher interactions, composite-operator renormalization in an unrestricted
ultraviolet extension, and any independent effective-field-theory cutoff are
logically distinct questions.  In addition, the full nonlinear
${\cal R}_{\rm eff}$ introduced in Sec.~\ref{subsec:R-eff} opens the way to a
direct pre-bounce and near-bounce correlation-function analysis in which the
time-derivative, gradient, inverse-Laplacian, and shift contributions can no
longer be discarded.

Within the retained ABS-type nonlinear-conversion diagnostic, the
bounce-state estimates are small at the five tested momenta in the specified
finite window. The remote-state signed results are cancellation sensitive and
refer to a scale-adapted window. Further checks of the numerical integrals,
omitted interaction terms, and primordial template are required before drawing
conclusions about the full interacting theory or the precision of the CMB
projection.

\label{v5removed:41}
\begin{acknowledgments}

J.S. and R.P. acknowledge the generous support of Baylor University through the Department of Physics and Astronomy graduate program. A.W. is partially supported by the U.S. National Science Foundation under Grant No.~PHY-2308845. 
We also acknowledge the use of ChatGPT (OpenAI) as a computational aid in checking mathematical derivations, assisting with numerical calculations, and developing codes for some of the figures. The authors take full responsibility for the scientific content of this work.

\end{acknowledgments}

\appendix
\section{Useful algebraic relations}
\renewcommand{\theequation}{A.\arabic{equation}}
\setcounter{equation}{0}
\lb{app:relations}

The relations used repeatedly in converting from $(a,\pi_a)$ to
$(v,\Omega)$ are
\bq
v=a^3,
 \quad
\pi_a=-\frac{6\Omega}{\kappa\gamma a},
\lb{eq:app-basic}
\eq
which imply
\bq
\frac{1}{\pi_a}
=-\frac{\kappa\gamma a}{6\Omega},
 \quad
\frac{1}{\pi_a^2}
=\frac{\kappa^2\gamma^2a^2}{36\Omega^2},
\lb{eq:app-pi-inv12}
\eq
and
\bq
\frac{1}{\pi_a^3}
=-\frac{\kappa^3\gamma^3a^3}{216\Omega^3}.
\lb{eq:app-pi-inv3}
\eq
For the hybrid momentum combination,
\bq
\delta\bar p_\phi
=a\left(\pi_\nu+\frac{\kappa\pi_a}{6a}\nu\right)
=a\pi_\nu-\frac{\Omega}{\gamma a}\nu
=v^{1/3}{\cal P}_\nu.
\lb{eq:app-pbar}
\eq

\section{Canonical completion on the full phase space}
\renewcommand{\theequation}{B.\arabic{equation}}
\setcounter{equation}{0}
\label{app:canonical-completion}

In this appendix we verify explicitly that the background-dependent
transformation from
$(\delta\phi,\delta p_\phi^{\rm old})$ to $(\nu,\pi_\nu)$ can be
completed canonically on the full phase space without generating any
additional cubic contribution to the perturbative Hamiltonian.

For a fiducial cell of coordinate volume $V_0$, we define
\begin{equation}
 \langle F\rangle
 \equiv
 \frac{1}{V_0}
 \int_{V_0}d^3x\,F .
 \label{eq:cg-average-app}
\end{equation}
The perturbative transformation can be written as
\begin{equation}
 \delta\phi=\frac{\nu}{a},
  \quad
 \delta p_\phi^{\rm old}
 =
 a\pi_\nu+a{\cal K}\,\nu ,
 \label{eq:app-full-map}
\end{equation}
where
\begin{equation}
 {\cal K}
 \equiv
 \frac{\kappa\pi_a}{6a}
 +
 \frac{3p_\phi^2}{a^3\pi_a}.
 \label{eq:app-D-def}
\end{equation}

Notice that
\begin{equation}
 \delta\bar p_\phi
 =
 \delta p_\phi^{\rm old}
 -
 \frac{3p_\phi^2}{a\pi_a}\delta\phi
 =
 a\left(
 \pi_\nu+\frac{\kappa\pi_a}{6a}\nu
 \right),
 \label{eq:app-pbar-check}
\end{equation}
as used in the main text.

The symplectic potential before the transformation is
\begin{equation}
 \Theta
 =
 V_0\pi_a\,da
 +
 V_0p_\phi\,d\phi
 +
 \int_{V_0}d^3x\,
 \delta p_\phi^{\rm old}\,
 d(\delta\phi).
 \label{eq:app-symplectic-old}
\end{equation}
Substituting Eq.~(\ref{eq:app-full-map}), integrating an exact
differential, and retaining terms through quadratic order in the
perturbations gives
\begin{align}
 \Theta
 ={}&
 V_0\pi_a\,da
 +
 V_0p_\phi\,d\phi
 +
 \int_{V_0}d^3x\,\pi_\nu\,d\nu
 \nonumber\\
 &+
 V_0
 \left[
 \Delta\pi_a\,da
 -
 \Delta a\,d\pi_a
 +
 \Delta p_\phi\,d\phi
 -
 \Delta\phi\,dp_\phi
 \right]\nonumber\\
 &
 +d{\cal F}_0
 +{\cal O}(\epsilon^4),
 \label{eq:app-symplectic-expanded}
\end{align}
where $\epsilon$ denotes the perturbative amplitude. The exact
differential before completing the homogeneous variables is
\begin{equation}
 {\cal F}_0=\frac{V_0}{2}{\cal K}\langle\nu^2\rangle .
 \label{eq:app-generating-F0}
\end{equation}

The required quadratic redefinitions of the homogeneous variables are
\begin{equation}
 {
 \Delta a
 =
 \left(
 \frac{\kappa}{12a}
 -
 \frac{3p_\phi^2}{2a^3\pi_a^2}
 \right)
 \langle\nu^2\rangle ,
 }
 \label{eq:app-Delta-a}
\end{equation}
\begin{equation}
 {
 \Delta\phi
 =
 \frac{3p_\phi}{a^3\pi_a}
 \langle\nu^2\rangle ,
 }
 \label{eq:app-Delta-phi}
\end{equation}
\begin{equation}
 {
 \Delta p_\phi=0 ,
 }
 \label{eq:app-Delta-pphi}
\end{equation}
and
\begin{equation}
 {
 \Delta\pi_a
 =
 -\frac{1}{a}\langle\nu\pi_\nu\rangle
 -\frac{\kappa\pi_a}{12a^2}\langle\nu^2\rangle
 +\frac{3p_\phi^2}{2a^4\pi_a}
 \langle\nu^2\rangle .
 }
 \label{eq:app-Delta-pia}
\end{equation}

Introducing the completed homogeneous variables
\bqn %egin{equation}
 \widetilde a&=&a+\Delta a,
  \quad
 \widetilde\pi_a=\pi_a+\Delta\pi_a,
  \quad
 \widetilde\phi=\phi+\Delta\phi,
  \nb\\
 \widetilde p_\phi&=&p_\phi+\Delta p_\phi,
 \label{eq:app-completed-background}
\eqn %nd{equation}
we obtain
\begin{align}
 \Theta
 ={}&
 V_0\widetilde\pi_a\,d\widetilde a
 +
 V_0\widetilde p_\phi\,d\widetilde\phi
 +
 \int_{V_0}d^3x\,\pi_\nu\,d\nu
 \nonumber\\
 &+
 d{\cal F}_{\rm can}
 +{\cal O}(\epsilon^4).
 \label{eq:app-symplectic-canonical}
\end{align}
The remaining exact differential is determined by
\begin{equation}
 {\cal F}_{\rm can}
 ={\cal F}_0-V_0(\pi_a\Delta a+p_\phi\Delta\phi)
 ={\cal O}(\epsilon^4).
 \label{eq:app-generating-Fcan}
\end{equation}
Indeed, ${\cal K}$ is homogeneous of degree one in $(\pi_a,p_\phi)$,
so Euler's identity gives
$\pi_a\partial_{\pi_a}{\cal K}+p_\phi\partial_{p_\phi}{\cal K}
={\cal K}$. Thus the displayed endpoint term cancels through quadratic
order after completing the homogeneous variables. The transformation is
canonical through the order required for the stated minimal cubic extension.
This cancellation concerns this particular canonical completion; boundary
terms generated by other integrations by parts, and initial-state
transformations between different canonical representations, require their
own accounting.

It remains to check how the quadratic background redefinitions affect
the perturbative expansion.  Denote collectively the homogeneous
variables by $W^A$ and the perturbative variables by $V$.  Since
\begin{equation}
 \Delta W^A={\cal O}(\epsilon^2),
 \label{eq:app-background-order}
\end{equation}
and
\begin{equation}
 W^A=\widetilde W^A-\Delta W^A,
\end{equation}
the background Hamiltonian transforms according to
\begin{equation}
 H^{(0)}(W)
 =
 H^{(0)}(\widetilde W)
 -
 \Delta W^A
 \frac{\partial H^{(0)}}{\partial W^A}
 \bigg|_{\widetilde W}
 +
 {\cal O}(\epsilon^4).
 \label{eq:app-H0-transform}
\end{equation}
The second term in Eq.~(\ref{eq:app-H0-transform}) is of quadratic
perturbative order and is therefore part of the completed quadratic
Hamiltonian.

For the second-order Hamiltonian, we find
\begin{equation}
 H^{(2)}(W,V)
 =
 H^{(2)}(\widetilde W,V)
 +
 {\cal O}(\epsilon^4),
 \label{eq:app-H2-transform}
\end{equation}
because
$\Delta W^A\,\partial_A H^{(2)}
={\cal O}(\epsilon^4)$.
Likewise, we find
\begin{equation}
 H^{(3)}(W,V)
 =
 H^{(3)}(\widetilde W,V)
 +
 {\cal O}(\epsilon^5).
 \label{eq:app-H3-transform}
\end{equation}
Therefore the quadratic background shifts required for the canonical
completion do not generate any additional cubic contribution.

The only other possible source of such a contribution would be a term
of the schematic form
$\Delta W^{(2)}\,\partial_W H^{(1)}$.
However, after imposing the spatially flat gauge and solving the
first-order scalar and vector constraints, the reduced perturbative
Hamiltonian starts at second order,
\begin{equation}
 H_{\rm pert}
 =
 H^{(2)}+H^{(3)}+\cdots .
 \label{eq:app-no-H1}
\end{equation}
Hence no such cubic term is present.

We conclude that, within the minimal nonlinear extension in which
$\nu=a\,\delta\phi$ is kept linear, the full canonical completion leaves
the explicit cubic Hamiltonian unchanged:
\begin{equation}
 {
 \widetilde H^{(3)}
 =
 H^{(3)}
 \left[
 \delta\phi=\frac{\nu}{a},
 \quad
 \delta\bar p_\phi
 =
 a\left(
 \pi_\nu+\frac{\kappa\pi_a}{6a}\nu
 \right)
 \right].
 }
 \label{eq:app-H3-final}
\end{equation}
All homogeneous variables appearing on the right-hand side of
Eq.~(\ref{eq:app-H3-final}) are understood as the canonically completed
variables, with the tildes suppressed for notational simplicity.

\section{Independent numerical validation and a fixed-window remote-state test}
\renewcommand{\theequation}{C.\arabic{equation}}
\setcounter{equation}{0}
\label{app:independent-validation}
This appendix reports independent implementations of selected background,
mode, vertex, and convolution equations.  Their purpose is to distinguish
algebraic consistency, numerical stability on a specified finite domain,
and the physical scope of the retained diagnostic.  Agreement of two
implementations sharing a Hamiltonian does not independently establish that
Hamiltonian or control contributions omitted from
Eq.~(\ref{eq:two-point-power-counting}).  The original numerical tables are
retained: the explicitly specified tests below do not reconstruct every
initial-state, averaging, and momentum-window convention underlying them.

\label{v5removed:44}

\subsection{Conventions and controlled benchmarks}
\label{app:validation-conventions}

The independent calculations set $G=\hbar=c=1$, with the unreduced Planck
mass $m_{\rm P}=1$ and $\kappa=8\pi$.  The representative inputs are
\begin{align}
 \gamma&=0.2375, &
 \lambda^2&=4\sqrt{3}\pi\gamma\ell_{\rm P}^2,\nonumber\\
 m_\star&=3.08\times10^{-6}m_{\rm P}, &
 \phi_B&=-1.09m_{\rm P},
 \label{eq:independent-inputs}
\end{align}
with $a_B=1$ and positive $\dot\phi_B$.  
% \textcolor{blue}{The area-gap choice is an explicit convention for these tests.  It gives
% $\rho_B\simeq0.09687888m_{\rm P}^4$ and $k_B\simeq1.6015113m_{\rm P}$, rather than the original adopted scale in}
% Eq.~(\ref{eq:kB-num}).  
These two scale choices are kept distinct when
comparing rapidly oscillating quantities.  On the expanding branch the
Hamiltonian-constraint root is used after the early evolution; subsequent
constraint residuals are therefore algebraically enforced, rather than an
independent conservation test.

The in-in sign and contraction factor are checked with a separate finite-time
oscillator, $H=\omega(n+1/2)+gq^3$, initialized in the free vacuum.  The
connected cumulant, rather than the raw third moment containing a tadpole,
satisfies at first order
\begin{equation}
 \kappa_3(T)=-\frac{g}{2\omega^4}
 \bigl[1-\cos(3\omega T)\bigr]+O(g^3).
 \label{eq:independent-oscillator}
\end{equation}
Finite-Fock-space evolution and coupling extrapolation reproduce this
coefficient to approximately $1.3\times10^{-12}$ relatively in the tested
cases.  This is a finite-time weak-coupling benchmark, not a statement about
the ground state of an unbounded cubic potential.

A second benchmark uses a separate classical GR background with constant
$\epsilon=-\dot H/H^2$ and analytic Bunch--Davies Hankel modes.  With the
negative $f_{\rm NL}$ convention of Eq.~(\ref{eq:fNL-def}), the leading
slow-roll equilateral and exact power-law soft-limit references are
\begin{align}
 f_{\rm NL}^{\rm equil}&=-\frac{55}{36}\epsilon+O(\epsilon^2),\nonumber\\
 f_{\rm NL}^{\rm soft}&=-\frac{5\epsilon}{6(1-\epsilon)}.
 \label{eq:independent-GR-limits}
\end{align}
The finite-momentum shape agrees with the leading result of
Ref.~\cite{Maldacena:2002vr} under a small-$\epsilon$ extrapolation at the
$10^{-5}$ level or better for the three tested shapes.  At long-to-short
momentum ratio $5\times10^{-4}$ and a refined late surface, the soft-limit
residual is about $4.5\times10^{-7}$ in the tested cases.  These residuals
include finite-momentum and finite-final-time effects.  The GR vacuum
contour and soft relation are not imposed on the bounce or remote candidate
states.  An independent symplectic check also supports the minimal canonical
completion in Appendix~\ref{app:canonical-completion}; it does not establish
uniqueness of the full quantum nonlinear transformation.
 A separate finite-dimensional unitary-transformation test of the cubic
oscillator reproduces the connected evolution with consistently transformed
initial data; resetting the free Gaussian changes the result. This is a
boundary/state-bookkeeping benchmark, not evidence for an unidentified
boundary term in the mLQC-I Hamiltonian.

\subsection{Exact radial separation and oscillatory time quadrature}
\label{app:validation-quadrature}

Expanding the scalar products and inverse squared momenta in
Eq.~(\ref{eq:Vexact}) gives a finite separable representation,
\begin{equation}
 \mathscr V_{kpq}
 =\sum_{\ell=1}^{49}c_\ell(\eta)
 \prod_{j=1}^{3}k_j^{2e_{\ell j}}X_{\ell j,k_j}(\eta),
 \label{eq:independent-49terms}
\end{equation}
where $(k_1,k_2,k_3)=(k,p,q)$, each $X$ is one of the mode combinations in
Eq.~(\ref{eq:upy}), and $e_{\ell j}\in\{-1,0,1,2\}$.  The coefficients
absorb all background factors.  Random-triangle algebra checks of this
49-term expansion agree with the compact vertex to about
$2.7\times10^{-13}$ relatively in the tested sample.

For example, an internal contraction is organized as
\begin{equation}
 \int_{p_-}^{p_+}\dd p\,F_\ell(p,\eta)
 \int_{q_-(k,p)}^{q_+(k,p)}\dd q\,G_\ell(q,\eta),
 \label{eq:independent-radial-primitive}
\end{equation}
with the final external modes and physical $pq$ measure included in the
factors.  Cumulative one-dimensional primitives retain the actual moving
triangle limits.  The algebraic separation is exact; interpolation and
quadrature of its factors are not.  It also does not turn an integral of the
absolute bispectrum into a product of absolute vertex integrals.

The numerical implementation evolves $r_k=u_k/a$ and the corresponding
rescaled momentum and shift combinations.  As a separate reduction check,
writing $S_{kpq}=B_\nu(k,p,q)/a_f^3$ gives
\begin{align}
 \delta_B={}&-\frac{C_{\calR}H_f}
 {2\pi^2k\dot\phi_f|r_{k,f}|^2}
 \int\!\dd p\,\dd q\,pq\,S_{kpq},\nonumber\\
 \delta_W={}&\frac{C_{\calR}^2H_f^2}
 {2\pi^2k\dot\phi_f^2|r_{k,f}|^2}
 \int\!\dd p\,\dd q\,pq\,|r_{p,f}|^2|r_{q,f}|^2.
 \label{eq:independent-reduced-kernels}
\end{align}
Both integrals use the same finite triangle domain.  Independent reductions
of stored kernels check the measure and prefactors.  Direct triangle and
cumulative-radial implementations agree on local square and moving-boundary
panels and on a common full-domain coarse lattice.  Agreement on the same
undersampled lattice is not a continuum-convergence result.

Late and time-dependent modes are represented locally by complex polynomial
envelopes multiplying explicit sine and cosine carriers.  Separate momentum
nodes validate these representations.  On the final-time mode test, the worst
sampled panel RMS-relative residual for degree-ten envelopes is about
$4.9\times10^{-10}$.  Fine radial grids are evaluated from these interpolants;
they are not newly propagated modes at every fine-grid momentum.  Composite
Simpson rules, with a final three-eighths block when required, preserve the
aligned triangle endpoints.  Their spatial error is tested separately.

Ordinary time refinement remains inefficient on the rapidly oscillating
post-bounce interval.  For the control window below, candidate time carriers
are derived from the domain vertices through
$|\pm k\pm p_v\pm q_v|$.  In Planck units they are
\begin{equation}
 \{0,0.1,0.3,0.5,0.6,0.7,1.5,1.8,2.1,2.4\}.
 \label{eq:independent-time-carriers}
\end{equation}
Their sufficiency for the radially integrated kernel is a numerical
hypothesis, not an exact single-harmonic identity.  On each local interval,
slow polynomial envelopes in $N$ are fitted to these carriers in $\eta(N)$.
The time contraction keeps the Jacobian explicitly,
\begin{equation}
 \int g(N)\,\dd N
 =\int aH\,g\bigl[N(\eta)\bigr]\,\dd\eta.
 \label{eq:independent-time-Jacobian}
\end{equation}
Sine/cosine-weighted quadrature integrates each carrier; the zero-frequency
term is integrated directly.  Synthetic weighted-quadrature checks agree
with direct high-order integration to about $7.2\times10^{-14}$ absolute.
Physical short-interval checks on two pilot panels also agree, with suitable
envelope degrees, at approximately $10^{-15}$ absolute or better.  Broad
fits and rank-deficient higher-degree fits that fail validation are rejected.

The critical $N\in[2,6]$ interval is divided into sixteen panels of width
$0.25$.  Each uses 129 Chebyshev training times and 32 separate validation
times, giving 2561 unique physical time nodes.  Acceptance requires full
column rank, condition number below $10^{10}$, sampled relative RMS below
$10^{-4}$, maximum sampled absolute residual times panel width below
$10^{-10}$, and stability against an independently eligible higher degree.
Ordinary polynomial quadrature is retained on the four $N\in[2,3]$ panels
where it already resolves the oscillations; the twelve $N\in[3,6]$ panels use
phase-aware quadrature.  All sixteen pass these gates and a final test on
128 additional time nodes not used to fit or select the models.

For the $t_i=-25t_{\rm P}$ candidate, the $N\in[2,6]$ contribution to
$\delta_B$ is approximately $2.585\times10^{-8}$.  The sums of
maximum-sampled-residual-times-width indicators from the selection and final
tests are about $6.7\times10^{-11}$ and $5.4\times10^{-11}$; the summed
higher-degree integral changes are about $8.0\times10^{-13}$.  Matched
radial-grid checks give a separate sampled indicator of approximately
$5.2\times10^{-10}$.  Tightening the mode tolerance from $10^{-12}$ to
$2\times10^{-13}$ on the final test nodes changes physical radial derivatives
by at most about $1.3\times10^{-16}$.  None of these sampled indicators or
internal quadrature estimates is a certified total physical error bound.

\subsection{Conditional signed result and finite initial-time robustness}
\label{app:validation-remote-result}

The fixed-window remote candidate calculation uses
\begin{equation}
 k=0.3m_{\rm P},\quad
 0.05\leq\frac{p}{m_{\rm P}},\frac{q}{m_{\rm P}}\leq1.05,
 \quad N_f=18,
 \label{eq:independent-control-window}
\end{equation}
with the triangle inequalities imposed throughout.  This is neither the
scale-adapted window of Eq.~(\ref{eq:selfsimilar-window}) nor the enlarged
five-point scan.  Combining the phase-aware $N\in[2,6]$ integral with the
earlier history and small later-time tail gives, for the specified
$t_i=-25t_{\rm P}$ candidate,
\begin{align}
 \delta_B&\simeq3.4485\times10^{-4},\nonumber\\
 \delta_W&\simeq9.5469\times10^{-10},\qquad
 \Xi\simeq3.45\times10^{-4}.
 \label{eq:independent-remote-Xi}
\end{align}
Only the rounded $\Xi$ should be interpreted as the present conditional
numerical estimate.  The dominant contribution comes from the contracting
history, particularly $t/t_{\rm P}\in[-7,-4]$.  Its degree-ten to degree-twelve
envelope comparison is stable, while a separate radial refinement from
80001 to 160001 points changes that dominant panel by about
$1.9\times10^{-9}$ absolutely.  The smaller $N\in[6,18]$ tail retains its
separate temporal-resolution qualification; it is not newly phase-validated
by the $N\in[2,6]$ test.  A matched final-surface comparison is reported
in Appendix~\ref{app:validation-final-surface}; window refinements,
background errors, and interpolation amplification under cancellation
remain separate uncertainties.

Fresh starts at $t_i/t_{\rm P}=-20,-25,-30$ provide a full signed-convolution
comparison, not merely a comparison of late powers or equilateral
bispectra.  To isolate the starting prescription, the early radial grid is
matched at 40001 points in all three runs; the critical later radial grid is
20001 points with separate 40001-point controls.  This matched reference
therefore differs slightly from the higher-resolution early integral used
in Eq.~(\ref{eq:independent-remote-Xi}).  It is not a new physical shift.

\begin{table}[!tbp]

 \caption{Finite-start robustness of the independent remote candidate on
 Eq.~(\ref{eq:independent-control-window}).  Percentage differences use the
 matched $t_i=-25t_{\rm P}$ estimate as denominator.  They describe differences
 of numerical estimates, not certified total accuracies.}
 \label{tab:independent-initial-time}
 \begin{ruledtabular}
 \begin{tabular}{ccc}
 $t_i/t_{\rm P}$ & $100(\Xi_i/\Xi_{-25}-1)$ & passed panels\\
 \hline
 $-20$ & $-0.00059\%$ & $16/16$\\
 $-25$ & $0$ & $16/16$\\
 $-30$ & $-0.0120\%$ & $16/16$
 \end{tabular}
 \end{ruledtabular}
\end{table}

All three values round to $\Xi\simeq3.45\times10^{-4}$.  Both new runs also
pass their independent final time-node tests.  Linear reuse of the stored
histories is checked through the canonical identity
$u_k^{\rm new}=\alpha_k u_k^{\rm ref}+\beta_k u_k^{{\rm ref}*}$ at a common
time, with fresh direct forward evolutions providing a separate comparison.
The maximum final-mode discrepancy divided by the mode RMS is below
$4.1\times10^{-10}$ in these tests.  On the dominant early panel, replacing
the transported final modes with direct ones changes the integral by less
than $1.3\times10^{-13}$.  Refining its matched radial grid from 40001 to
80001 points changes the state-to-reference differences by at most about
$5.1\times10^{-12}$, much less than the observed starting-prescription shifts.

Each fresh run imposes zero initial connected three-point function at its
own surface.  Thus both free-state data and nonlinear initial conditions
vary.  A separate test holds the free reference state fixed and changes only
the lower interaction limit with a fresh zero primitive; its shifts from the
$-25t_{\rm P}$ reference are approximately $-1.106\times10^{-8}$ and
$+1.106\times10^{-8}$ for the $-20t_{\rm P}$ and $-30t_{\rm P}$ cutoffs.
These effects partially compensate free-state changes in one of the fresh
initializations.  Consistent transport of one interacting state would retain
its accumulated primitive and is a different operation.

The results support a small retained signed correction and finite-start
robustness for the specified candidate and domain.  They do not establish
an infinite-past state, a uniform external-momentum or window bound, the
complete same-order two-point correction, an error bound on discarded
conversion terms, or a validated primordial-to-CMB projection.  In
particular, Eq.~(\ref{eq:independent-remote-Xi}) is not a replacement for any entry of Table~\ref{tab:selfsimilar-Xi}. 

\subsection{Dependence on the final evaluation surface}
\label{app:validation-final-surface}

A separate matched calculation holds the candidate state at
$t_i=-25t_{\rm P}$, the external momentum $k=0.3m_{\rm P}$, and the
control window in Eq.~(\ref{eq:independent-control-window}) fixed, and
varies only the final surface.  The interaction history is integrated
through each final surface without resetting its accumulated primitive.
The early, critical later-time, and Wick radial grids contain 40001,
20001, and 160001 points, respectively.  For $N_F=18,19,20$ the retained
diagnostic is, respectively,
\begin{equation}
 \Xi\simeq(3.45,\,3.47,\,3.50)\times10^{-4}.
 \label{eq:independent-final-surface}
\end{equation}
All sixteen critical time panels and their independent unused time-node
checks pass at each final surface.  These are sampled numerical checks,
not certified bounds on the full integration error.

The matched unrounded estimates increase by $0.734\%$ and $1.479\%$
relative to $N_F=18$.  This drift closely tracks the local conversion
coefficient: $C_{\cal R}=0.00363282,0.00365949,0.00368655$ on the three
surfaces.  In contrast, the leading external curvature power changes by
only about $8.60\times10^{-6}$ fractionally from $N_F=18$ to $20$.
The ratio $\delta_B/C_{\cal R}$ changes by about
$7.57\times10^{-7}$ fractionally.  This ratio diagnoses the scaling;
it does not replace or redefine $\Xi$.  Tightening the mode-solver
tolerance from $10^{-12}$ to $2\times10^{-13}$ changes the dominant
early-panel contribution at $N_F=20$ by only about $5.3\times10^{-14}$.

The retained correction therefore remains small in this test, but is
not strictly independent of the late evaluation surface even though
the linear curvature power is nearly frozen.  Because the other
same-order contributions and omitted conversion terms have not been
computed, this drift cannot by itself establish the evolution of the
complete physical two-point function.  The fixed-$N_F=18$ conditional
estimate above remains applicable on its stated surface; a stronger
claim of late-time invariance is not supported by this comparison.

\label{v5removed:49}

\section{Details of the RBF interpolation}
\renewcommand{\theequation}{D.\arabic{equation}}
\setcounter{equation}{0}
\label{app:RBF-complete}

For completeness, we summarize the radial-basis interpolation used in
Sec.~VI.E.  For each state $s=B,R$ we define the $N_{\rm node}=148$ nodes
\bq
\lb{eqc.1}
 {\bf y}_j=(x_j,U_j,V_j),\quad
 x_j=\ln\!\left(\frac{k_{t,j}}{8m_{\rm P}}\right),
\eq
with $U_j$ and $V_j$ defined in Eq.~(\ref{eq:UV-shape}). The fitted data are
\bq
\lb{eqc.2}
 d_{A,j}=\ln S_{f,j},
 \qquad
 d_{\varphi,j}=\Delta\varphi_j ,
\eq
These are the logarithmic amplitude ratio and relative phase shift used as
the two interpolation targets, respectively.

The unconstrained radial-basis functions in Eq.~(\ref{eq:RBF-template}) are
reconstructed from
\bq
\lb{eqc.3}
 \begin{pmatrix}
 K+\lambda_X^{(s)} I & P\\
 P^T & 0
 \end{pmatrix}
 \begin{pmatrix}
 {\bf w}_X^{(s)}\\ {\bf c}_X^{(s)}
 \end{pmatrix}
 =
 \begin{pmatrix}
 {\bf d}_X^{(s)}\\ 0
 \end{pmatrix},
\eq
where
\bq
\lb{eqc.4}
 K_{ij}=\Phi(|{\bf y}_i-{\bf y}_j|),\quad
 \Phi(\rho)=\rho^2\ln\rho,\quad \Phi(0)=0,
\eq
and
\bq
\lb{eqc.5}
 P_j=(1,x_j,U_j,V_j).
\eq
Thus, the polynomial completion is linear,
\bq
\lb{eqc.6}
 p_X^{(s)}({\bf y})=
 c_{X0}^{(s)}+c_{Xx}^{(s)}x+c_{XU}^{(s)}U+c_{XV}^{(s)}V,
 \quad X=A,\varphi .
\eq
The smoothing parameters and polynomial coefficients are given in
Table~\ref{tab:RBF-poly-coeff}.
\begin{table}[ht!]
\caption{Smoothing parameters and polynomial-completion coefficients for the
RBF representation.}
\label{tab:RBF-poly-coeff}
\centering
\resizebox{\columnwidth}{!}{%
\begin{tabular}{ccccccc}
\hline\hline
state & target & $\lambda$ & $c_0$ & $c_x$ & $c_U$ & $c_V$\\
\hline
$B$ & $\widetilde A$ & $10^{-5}$ & $-0.012013731$ & $0.078259715$ & $-0.96631451$ & $9.3018474$\\
$B$ & $\widetilde\Phi$ & $10^{-5}$ & $0.25810551$ & $0.16018171$ & $1.3597425$ & $7.270179$\\
$R$ & $\widetilde A$ & $10^{-5}$ & $2.7558132$ & $0.19096987$ & $-12.918395$ & $-34.459592$\\
$R$ & $\widetilde\Phi$ & $10^{-4}$ & $0.26669288$ & $-0.093977977$ & $0.39972875$ & $1.947794$\\
\hline\hline
\end{tabular}%
}
\end{table}

The equilateral normalization is imposed by
\bqn
\lb{eqc.7}
&& \ln S_f^{(s)}(x,U,V)
 =\widetilde A^{(s)}(x,U,V)-\widetilde A^{(s)}(x,0,0), ~~~~
\\
\lb{eqc.8}
&& \Delta\varphi^{(s)}(x,U,V)
 =\widetilde\Phi^{(s)}(x,U,V)-\widetilde\Phi^{(s)}(x,0,0). ~~~~
\eqn
The node coordinates and fitted targets are those obtained from the 37
triangle shapes at each of the four reference values of $k_t$ described in
Sec.~VI.E.  Equations~(\ref{eqc.1})--(\ref{eqc.8}), together with
Table~\ref{tab:RBF-poly-coeff}, summarize the RBF construction used in the numerical analysis.

\bibliography{BibNVar}

\end{document}